\documentclass[useAMS,usenatbib,usegraphicx]{mn2e}
\usepackage{amsmath}

\newcommand{\lapprox}{\stackrel{<}{\scriptstyle\sim}}
\newcommand{\gapprox}{\stackrel{>}{\scriptstyle\sim}}

\title[Dust grain dynamics in  C-Type shock waves  in molecular clouds]
  {Dust grain dynamics in C-Type shock waves in molecular clouds}
\author[J. F. Chapman and M. Wardle]{J. F.~Chapman$^{1,2}$\thanks{E-mail:
jchapman@physics.mq.edu.au} and Mark Wardle$^{2}$\thanks{E-mail:
wardle@physics.mq.edu.au}\\
  $^1$ School of Physics, University of Sydney, NSW 2006,  Australia\\
  $^2$ Department of Physics,  Macquarie University, Sydney, NSW 2109, Australia}

\begin{document}


\date{Submitted to MNRAS 29 July 2005}
\pagerange{\pageref{firstpage}--\pageref{lastpage}}\pubyear{0000}
\maketitle

\label{firstpage}

\begin{abstract}

We investigate the role and behaviour of dust grains in C-type MHD
shock waves in weakly ionized, dense molecular clouds.  The
structure of C-type shocks is largely determined by the coupling of
the charged species and the magnetic field. In weakly ionized
clouds,  charged dust grains enhance the energy and momentum
transfer between the magnetic field and the neutral fluid, and
dominate the neutral collisional heating rate. New shock models are
developed here for steady oblique C-type shock structures with shock
speed $v_s=18\mathrm{km/s}$,   pre-shock number density
$n_H=10^5{\mathrm{cm}^{-3}}$, and a grain population represented by
either a single grain species or a MRN grain size distribution. The
grain size distribution is calculated using Gauss-Legendre weights
and the integrals over the continuous distribution of grain sizes
are converted to a  series of separate grain bins or `size classes'.
The dynamics of each grain size class various through the shock
front; smaller grains remain coupled to the magnetic field and
larger grains are partially decoupled from the magnetic field due to
collisions with the neutrals. The charges on the grains are allowed
to vary, via the sticking and re-releasing of electrons, increasing
with increasing electron temperature. The increase in grain charge
increases the coupling of the grains to the magnetic field, and
magnetic  field rotation out of the shock plane is suppressed.
MRN(mantles) and MRN(PAHs) distributions are also compared with the
standard MRN model. Increasing the grain sizes in the MRN(mantles)
model leads to an increase in the collisional heating of the
neutrals leading to hotter, thinner shock structures than those
using a standard  MRN distribution.  With the addition of PAHs, the
electron abundance is reduced and the grain charge is held constant,
resulting in less grain coupling to the magnetic field, and
substantial rotation of the magnetic field out of the shock plane.
The effects of varying the orientation of the pre-shock magnetic
field ${\bf B_0}$ with the shock normal, specified by the angle
$\theta$, are also considered. It is found that there are critical
values of $\theta$ below which the shock is no longer C-type and the
transition becomes C$^*$ or J-type. The degree of non-coplanarity of
the shock solution depends upon the grain model chosen,  as well as
the angle $\theta$.

\end{abstract}


\section{Introduction}

Interstellar shock waves play an integral part in the  chemical and
physical evolution of the interstellar medium. They are driven by
cloud-cloud collisions, stellar winds  and supernova explosions (e.g.
\citealt{MS79}).  They cause large increases in the local pressure
leading to both compression and heating of the gas. As a result, the
local increase in density leads to an  increase  in the rates at
which chemical reactions can occur. C-type shocks may be associated
with strong infrared emission lines \citep{DRD83}, and shock models
have been used to explain lines in the K-L region of Orion
\citep{DR82,CMH82}. C-type shocks may also be responsible for the
absorbtion lines along lines of sight, in particular of CH+, in
diffuse clouds \citep{FPH85,DK86a,DK86b,PEA86}.


The nature and structure of shock waves travelling through molecular
clouds  are strongly dependent upon the strength of the ambient
magnetic field ${\bf B}_0$ as well as the local ionization fraction.
When the magnetic  field strength is large and fractional ionization
low, a multifluid magnetohydrodynamic (MHD) description of the fluid
is usually required  \citep{M71}. In the multifluid treatment, there
are two speeds at which compressive disturbances may propagate. In
the charged component of the medium, disturbances can propagate at
the fast magnetosonic speed which is dependent upon both the
magnitude $|B_0|$ and the orientation of ${\bf B}_0$ with the local
shock normal. In the neutral fluid, compressive disturbances
propagate at the neutral sound speed.

The charged fluids gain energy and momentum from the magnetic field
${\bf B}$, and are coupled to the neutral fluid via collisions,
transferring  energy and momentum, acting to weakly couple the
neutrals to ${\bf B}$. If the fractional ionization rate is small
enough, leading to a low neutral-ion collision rate, the two fluids
can be thought of as separate and interpenetrating fluids
\citep{D80,DRD83}. The ions and electrons contribute only small
amounts of mass and pressure, however the magnetic fields are strong
enough to influence the shock dynamics.  If the shock speed is less
then the (fast) magnetosonic speeds of the charged fluids,  the
charged species are accelerated ahead of the neutrals, via ambipolar
diffusion in a magnetic precursor \citep{D80}. If the neutral
temperature remains low enough inside the shock and the neutral flow
is everywhere supersonic,  the shock  is C-type and all the
hydrodynamic variables remain continuous.

At low enough densities, a three fluid system is adequate with the
inclusion of neutrals, ions and electrons. The grains may be ignored
since their contribution to the  energy and momentum of the neutrals
is negligible. For densities
$n_{\mathrm{H}}\gapprox10^2\mathrm{cm^{-3}}$ the energy and momentum
transferred from the charged dust grains to the neutrals can no
longer be neglected (e.g., \citealt{DRD83}),  and in regions of high
density, $n_{\mathrm{H}}\sim 10^5\mathrm{cm^{-3}}$, and the grains
dominate the collisional heating of the neutrals. The grain
abundances  in the interstellar medium obey a size  distribution
(e.g., \citealt{MRN77,DL84,WD01}), and for large grain sizes the
friction between the grains and neutrals can become great enough
that the grains are decoupled from ${\bf B}$.
 In the opposite limit, smaller grains may remain tied to ${\bf B}$.
Dust grains also  have the potential to drift with respect to the
neutral fluid and move out of the shock plane, since grain-neutral
collisions can cause their partial decoupling from ${\bf B}$
 leading to non-coplanar shock structures.

For perpendicular shocks with ${\bf B}_0$ oriented perpendicular to
the local shock normal {\bf \^n}, changes in ${\bf B}$  and the
dynamic variables inside the shock are restricted to the shock plane
containing ${\bf B}_0$ and {\bf \^n} (e.g., \citealt{DRD83,FP03}).
When  ${\bf B}_0$ is oriented obliquely with the shock normal, there
is a  current parallel to the ${\bf B}$  component transverse to the
shock propagation direction. ${\bf B}_\perp$, the ${\bf B}$
 component perpendicular to {\bf \^n}, can rotate out of the shock plane inside the shock before returning
to the shock plane downstream. \citet{WD87} demonstrated that by
changing  only the transverse component of ${\bf B}$, very different
hydrodynamic shock structures are found.

\citet{PH94} investigated the rotation of ${\bf B}$  in fast
steady shock structures with the inclusion of one grain species.
This involved integration of the shock fluid conservation equations
from upstream to downstream which proved difficult for shock
speeds $\gapprox 3 \mathrm{km s^{-1}}$. \citet{PH94}
concluded that steady fast shock solutions do not exist for these
conditions. \citet{W98} showed that these solutions were in fact intermediate,
not fast, solutions  which do not exist for shock speeds above a
critical value. \citet{W98} presented magnetic field ($B_x$ versus
$B_y$) phase space topologies, demonstrating that the shock solution
belongs to a one-parameter family. Integration from upstream becomes complicated since
the phase space trajectories can diverge due to finite numerical
precision, and end up on neighbouring intermediate solutions which
can become unphysical. Integrating  from the downstream state ensures that the correct
fast solution trajectory is followed.

In this paper we investigate the non-coplanar nature of C-type
shocks with the inclusion of a grain size distribution. The shock is
assumed steady state since the time scales over which the fluid
variables vary are longer then the time required to transverse the
shock. The methodology of \citet{W98} is extended to  include a
grain size distribution and the integration of the energy equation
allows for radiative cooling. The shock problem comprises of three
ordinary differential equations (ODEs) in $B_x$, $B_y$, and pressure
$P$, along with algebraic expressions for the neutral velocity
components, electric field components, and  charged species
properties. It is a two point boundary problem connecting the
upstream and downstream states, and integration proceeds from
downstream to upstream to ensure the  fast shock solution. With
radiative cooling the fluids  do not cool to their final temperature
until far downstream of the shock front, and integration over such
long time scales is infeasible (e.g., \citealt{D80}).  A shooting
integration method is implemented  which searches for an initial
state inside the cooling zone which, after integration towards
upstream, yields the  pre-shock state. Conditions inside  molecular
clouds are assumed, $n_H =10^5\mathrm{cm}^{-3}$, with shock speed
$v_s \sim 18~\mathrm{km~s^{-1}}$.

The derivation of the governing shock  equations are given in Section
\ref{fluidequations}, the shock jump conditions are
derived in Section \ref{jumpconditionsection}, and the heating and
cooling processes are discussed in Section
\ref{heatingandcoolingsection}. The  treatment of the  charged
species  are addressed in Section \ref{chargedspeciessection}, along
with the derivation for  both the electron and grain Hall parameters
and the electric field component $E_z'$, in Sections
\ref{sectionHallparameters} and  \ref{EZsection}, respectively.
The shock calculation and integration methods are explained
in Section \ref{shockcalculationsection} followed by  the results
 in Section \ref{resultssection}. A   discussion and summary
 are presented in Sections \ref{discussion} and \ref{summary}.

\section{Fluid Equations}\label{fluidequations}
The fluid is weakly ionized, and  the inertia and thermal pressures
of the charged species are neglected, as well as most processes that
create and destroy species.  There is  only the one exception, when
grain charging is used (equation (\ref{graincharge})) the electrons
are allowed to stick to the grains, and so electrons are removed
from,  or injected into, the electron fluid to conserve charge
density (equation (\ref{electronnumberdensity})). Each species is
characterized by a mass $m$, mass and number densities $\rho$ and
$n$, respectively, charge $Ze$ and  velocity ${\bf v}$. Charged
species are given a subscript of either $e$, $i$ or $g$, denoting
electrons, ions and grains, respectively. A subscript of $j$
represents any charged species. The pre-shock   medium  has mass and
number densities $\rho_0$ and  $n_0$, respectively, magnetic field
${\bf B}_0$, pressure $P_{0}$, temperature $T_0$ and the incoming
fluid has a speed $v_s$. The shock is assumed steady and
plane-parallel $\partial/\partial{t}=
\partial/\partial{x}=\partial/\partial{y}=0$ and the coordinate system
is defined in the shock frame with ${\bf B}_0$ lying in the $x-z$ plane making an angle  $\theta$
with the  $z$ axis. The incoming flow is parallel to the  $z$ axis.
The equations for a steady multifluid flow are then given by the
conservation of mass
\begin{eqnarray}
\frac{d \rho v_z}{dz} = 0, \label{max3}
\end{eqnarray}
momentum
\begin{eqnarray}
\rho v_z \frac{d{\bf v}}{dz} +\frac{d P}{dz}\hat{z}= \frac{{\bf
J}\times{\bf B}}{c}, \label{max4}
\end{eqnarray}
and energy
\begin{eqnarray}
\frac{d(Uv_z)}{dz} + P \frac{d v_z}{dz} = G - \Lambda,
\label{MHDpressure}
\end{eqnarray}
along with the following from  Maxwell's equations:
\begin{eqnarray}
\frac{d  E_{x,y}}{dz}=0 \label{max5}
\end{eqnarray}
and
\begin{eqnarray}
\frac{d B_z}{dz}=0. \label{max1}
\end{eqnarray}
Lastly, from Ampere's law:
\begin{eqnarray}
\frac{d{B_x}}{d{z}}&=& \frac{4\pi}{c}J_y,\label{current2}  \\
\frac{d{B_y}}{d{z}}&=& -\frac{4\pi}{c}J_x\label{current1}
\end{eqnarray}
and
\begin{eqnarray}
J_z=0. \label{current3}
\end{eqnarray}
$P$ and $U$ denote the neutral thermal pressure and internal
energy, $G$ and $\Lambda,$  the  heating and  cooling rates per unit
volume respectively, and ${\bf E}, {\bf B}$ and ${\bf J}$  are the
electric field, magnetic field, and current density, respectively, with
\begin{eqnarray}
P=\frac{\rho \mathrm{k} T }{m}. \label{pressure}
\end{eqnarray}
The internal energy $U$ is  the sum of the thermal kinetic energy and excitation energy of the
excited molecules
\begin{eqnarray}\label{internalone}
U = \left(\frac{3}{2}+\sum_l p_l\frac{E_l}{k T}\right)P
\end{eqnarray}
where the summation is over the electronic, rotational and
vibrational levels $l$, with corresponding probabilities $p_l$ and
energy $E_l$. In a $\mathrm{H}_2$ gas the  vibrationally excited
levels have negligible populations, and the rotational levels have a
thermalized distribution, and the sum  can be taken as  unity
\citep{W91, HC98} which is a good approximation in the hottest
part of the shock where the pressure becomes important. Equation (\ref{internalone}) becomes
\begin{eqnarray}\label{intenergy}
U=\frac{P}{\gamma-1}.
\end{eqnarray}
where $\gamma$  is the  adiabatic index with  $\gamma=7/5$.

 From mass conservation of the charged species
\begin{eqnarray}
 \frac{d}{dz} (\rho_jv_{zj})   = 0. \label{massionsandgrains}
\end{eqnarray}
The current density is given by
\begin{eqnarray}
{\bf J} = e \sum_{j} n_jZ_j{\bf v}_j,  = e \sum_{j} n_jZ_j({\bf v}_j-{\bf v} ), \label{currentdens}
\end{eqnarray}
assuming  charge neutrality
\begin{eqnarray}
\sum_{j} Z_jn_j=0.\label{max6}
\end{eqnarray}
$({\bf v}_j-{\bf v})$ is the drift velocity of a species $j$ with
respect to the neutrals.  The grains obey a continuous size
distribution implying  equation (\ref{max6}) contains an integral
over grain size.  However, the MRN size distribution is calculated
using Gauss-Legendre weights (Section \ref{grainsection}), and the
distribution is represented by a series of discrete grain size bins
each with a specified number density and charge. Each grain bin is
then considered a separate species $j$ in equation
(\ref{currentdens}).

The rotation of ${\bf B}_\perp (|{\bf B}_\perp|=\sqrt{B_x^2+B_y^2}$)
 inside the shock front is driven by
the Hall current along the ${\bf E' \times B}$ direction, and the
Pedersen current along the ${\bf E'}$ direction. The difference in
drift speeds between charged species drives the Hall current. In the
ambipolar diffusion limit all charged species are highly coupled to
${\bf B}$ and the Hall current vanishes. In this limit the magnetic
forces on the charged species dominates the neutral collisional drag
which is the case for the electrons and ions in molecular clouds.
When charged dust grains are present  the neutral-grain drag force
is non-negligible resulting in a non-zero Hall current  and there is
rotation of ${\bf B}_\perp$.
 \citet{WDNG99} demonstrated that for dense molecular
gas, the contribution  of dust grains to the Hall conductivity leads
to substantial changes in the dynamics of the gas as well as the
evolution of ${\bf B}$ .

Treating the charged particles as test particles, and neglecting their inertia,
 their drift through the neutrals is
\begin{eqnarray}\label{balanceforceequation}
n_jZ_je\left({\bf E} +\frac{{\bf v}_j}{c} \times {\bf B} \right)  + \gamma_j \rho_j \rho ({\bf
v}-{\bf v}_j) = 0,
\end{eqnarray}
which represents the balance between the electromagnetic forces
and  the collisional drag \citep{D80, SHU83}. The rate coefficient for elastic
scattering between particles of a species $j$ and the neutrals,
$<\sigma v>_j$, is related to $\gamma_j$  via
\begin{eqnarray}
\gamma_j = \frac{<\sigma v>_j}{m_j+m}. \label{gammaterm}
\end{eqnarray}
If $\sigma$ is not a function of $v$,  $<\sigma v>_j$ may be written as
$\sigma_j u_j$, where $u_j$ is the effective velocity  given by \citep{D86}
\begin{eqnarray}\label{effvel}
u_j = \left[ \varphi_j + |{\bf v}_j-{\bf v}|^2
\right]^{\frac{1}{2}},
\end{eqnarray}
where
\begin{eqnarray}
\varphi_j=\frac{128}{9\pi}\left(\frac{kT}{m_N} + \frac{k T_j}{m_j} \right). \label{const1Hall}
\end{eqnarray}
The treatment of $<\sigma v>_j$ for each charges species is discussed further below in Section
\ref{chargedspeciessection}.

Equation (\ref{balanceforceequation}) can be recast in terms of the electric field ${\bf E}'$ in
the frame co-moving with the neutrals,
\begin{eqnarray}
\frac{{\bf v}_j -{\bf v}}{c} = \beta_j\frac{({\bf B}\cdot{\bf E}'){\bf
B}}{B^3}+\frac{\beta_j^2}{1+\beta_j^2}\frac{{\bf E}'\times {\bf B}}{B^2} + \nonumber\\
\frac{\beta_j}{1+\beta_j^2}\frac{{\bf B}\times({\bf E}'\times {\bf
B})}{B^3}, \label{driftvel}
\end{eqnarray}
where
\begin{eqnarray}
{\bf E}'={\bf E} +{\bf v}\times{\bf B}/c \label{driftelec}
\end{eqnarray}
and $\beta_j$, the  Hall parameter for species $j$,
 \begin{eqnarray}
\beta_j = \frac{Z_jeB}{m_jc}\frac{1}{\gamma_j\rho} \label{Hallparm}
\end{eqnarray}
is the product of the gyrofrequency and the time-scale for momentum exchange with the neutral
fluid.

Integrating the mass conservation equations (\ref{max3}) and
(\ref{massionsandgrains}) along with charge neutrality (\ref{max6}) gives
(the subscript $k$ runs over the ions and
grains):
\begin{eqnarray}
\rho v_z= \rho_0 v_s \label{mass1}
\end{eqnarray}
\begin{eqnarray}
 \rho_k v_{zk}= \rho_{0k}
v_s \label{mass2}
\end{eqnarray}
\begin{eqnarray}\label{electronnumberdensity}
n_e=n_i+\sum_gZ_gn_g,
\end{eqnarray}
where a local equilibrium approximation is used for $Z_g$ (equation \ref{graincharge}).
Combining equations (\ref{max4}) with  equations
(\ref{current2})-(\ref{current3}), and integrating, gives
the neutral velocity components:
\begin{eqnarray}
v_x&=&\frac{1}{\rho_0 v_s}\left[\frac{B_z}{4\pi}(B_x-B_{0x})\right], \label{mom1}  \\
v_y&=&\frac{1}{\rho_0
v_s}\left[\frac{B_yB_z}{4\pi}\right],\label{mom2}
\end{eqnarray}
and
\begin{eqnarray}
v_z&=&\frac{1}{\rho_0 v_s}\left[P_0-P+ \rho_0
v_s^2+\frac{1}{8\pi}(B_{0x}^2-B_x^2-B_y^2)\right].\label{mom3}
\end{eqnarray}

Equations  (\ref{max4}) and (\ref{MHDpressure}) are combined, along
with (\ref{intenergy}), to construct an  ODE in $P$:
\begin{eqnarray}
\frac{dP}{dz} v_z\left(1-\frac{c_s^2}{v_z^2}\right)&=&(\gamma
-1)(G-\Lambda) \nonumber \\
&+&\frac{\gamma P}{4\pi\rho_0
v_s}\left[B_x\frac{dB_x}{dz}+B_y\frac{dB_y}{dz}\right],\label{odepressure}
\end{eqnarray}
where $c_s$ is the neutral sound speed
\begin{eqnarray}
c_{s}=\left(\frac{\gamma P}{\rho}\right)^{1/2}.
\end{eqnarray}
Equation (\ref{odepressure}) breaks down at the sonic point $v_z=c_s$.

Finally, from equations (\ref{max5}) and (\ref{max1}),
\begin{eqnarray}\label{electricfieldnonframe}
E_x=0, \,\,\,\,\,\,\,\,   E_y=-\frac{v_s}{c}B_{0x}
\end{eqnarray}
and
\begin{eqnarray}
B_z=B_{0z}=\cos \theta. \label{cond1}
\end{eqnarray}
Integrating equation (\ref{driftelec}), and using
(\ref{electricfieldnonframe});
\begin{eqnarray}
E_x'&=&(v_yB_z-v_zB_y)\frac{1}{c}\label{Electx}
\end{eqnarray}
and
\begin{eqnarray}
E_y'&=&(-v_sB_{0x} + v_zB_x-v_xB_z)\frac{1}{c}.\label{Electy}
\end{eqnarray}
Charge neutrality (equation \ref{max6}) is used to find  $E_z'$ (see
Section \ref{EZsection}).

The shock problem then comprises of three ODE's in $B_x, ~B_y$, and
$P$  equations  (\ref{current2}), (\ref{current1}), and
(\ref{odepressure}), respectively,  with the algebraic relations
(\ref{mass1}), (\ref{mom1})-(\ref{mom3}) for the neutral density and
neutral velocity components, respectively, along with equations
(\ref{cond1})-(\ref{Electy}) for $B_z$, $E_x'$ and $E_y'$. To
calculate the right hand side of the ODE's (\ref{current2}) and
(\ref{current1}), the  current density components $J_x$ and $J_y$
need to be known, which are dependent upon the drift velocities,
${\bf v}_j-{\bf v}$ (equation (\ref{currentdens})). Calculating
${\bf v}_j-{\bf v}$ from equation (\ref{driftvel}) requires
knowledge of the Hall parameters $\beta_j$, as well as the electric
field component $E_z'$. The treatment of $\beta_j$  and  $E_z'$ are
described below in Sections \ref{chargedspeciessection},
\ref{sectionHallparameters}, and \ref{EZsection}.   The heating and
cooling rates needed for the energy equation (\ref{odepressure}) are
discussed in  Section \ref{heatingandcoolingsection}.

\section{Jump Conditions}\label{jumpconditionsection}
The boundary conditions for  equations (\ref{current2}),
(\ref{current1}), and (\ref{odepressure})  are that the derivatives
vanish far upstream and downstream of the shock.  Shock solutions
begin and end at points where
\begin{eqnarray}
\frac{dB_x}{dz}=\frac{dB_y}{dz}=\frac{dP}{dz}=0.
\end{eqnarray}
From equations (\ref{current2})-(\ref{current3}),  ${\bf J}= 0$,
and equations (\ref{currentdens}) and (\ref{driftvel}) give ${\bf
E'}=0$ and ${\bf v}_j - {\bf v}=0$. All species are co-moving upstream
and downstream of the shock and  obey the same overall
jump conditions. Also Equation (\ref{odepressure}) gives
$G=\Lambda$. ${\bf E'}=0$,  so ${\bf E}=-{\bf v \times B}/c$ and by
equations (\ref{Electx}) and (\ref{Electy});
\begin{eqnarray}
v_yB_z-v_zB_y=0\label{jump1}
\end{eqnarray}
and
\begin{eqnarray}
v_zB_x-v_xB_z = v_sB_{0x}.\label{jump2}
\end{eqnarray}
Eliminating $v_y$ from equations (\ref{mom2}) and (\ref{jump2}) gives either
\begin{eqnarray}
B_y&=&0 \,\,\,\,\,\,\,\,\,\,\,\,\,\,\mathrm{or} \\ v_z&=&\frac{B_z^2}{4\pi\rho_0v_s}.
\end{eqnarray}
The second solution implies $v_z$ is equal to the intermediate
speed, resulting  in a rotational discontinuity  and  no
compression across the shock \citep{C76}. So taking $B_y= 0$ gives
the condition $v_y=0$  ahead of and behind the shock.

In a radiative shock, the downstream fluid radiates away the
energy transferred to it by the processes inside the shock.
 Cooling proceeds at a  rate $\Lambda(\rho, T)$
(Section \ref{heatingandcoolingsection}) over a post
shock cooling layer. Drift of the charged
particles inside the shock  lead to increases in thermal energy due to elastic scattering between
the charged and neutral components. The immediate post-shock
temperature is therefore much higher than the final post-shock
temperature after the fluid has radiated away the energy inside
the post-shock cooling layer.

The fluid cools inside the cooling layer at an almost  constant pressure as the density increases and
 fluid speed decreases.  The
post-shock fluid cools until it returns to the pre-shock temperature
$T_{0}$. Applying the isothermal jump conditions $T = T_{0}$  along with
equations (\ref{pressure}) and (\ref{mass1}) gives
\begin{eqnarray}
\frac{P}{P_{0}}=\frac{n}{n_0}= \frac{v_s}{v_z}. \label{isocond}
\end{eqnarray}

Equations (\ref{mom1}), (\ref{mom3}), (\ref{jump2}), and
(\ref{isocond})  give a cubic in $v_z/v_s$
 \begin{eqnarray}
 \frac{v_z^3}{v_s^3} + a_0\frac{v_z^2}{v_s^2}+a_1\frac{v_z}{v_s} +a_2
 =0,
 \label{isojump}
 \end{eqnarray}
 where
\begin{eqnarray}
a_0&=&-\frac{1}{2M_{A0}^2}(3\mu^2+1) -\frac{1}{\gamma M_{s0}^2},\\
a_1&=&\frac{1}{M_{A0}^2}\left[\frac{2\mu^2}{\gamma M_{s0}^2}
+\frac{\mu^2}{M_{A0}^2}-\frac{(1-\mu^2)}{2}\right],
\end{eqnarray}
and
\begin{eqnarray}
 a_2&=&-\frac{1}{\gamma}\frac{\mu^4}{M_{s0}^2M_{A0}^4}.
\end{eqnarray}
Only roots with $0<v_z/v_s<1$ are suitable for there to be
compression across the shock. Once the downstream  $v_{zd}$ is
calculated, the corresponding  $P$, $B_x$, and $v_x$  are easily
found. One could alternatively obtain a cubic in $B_x$ and firstly
solve for the downstream component $B_{xd}$ instead.

\section{Heating and Cooling Processes}\label{heatingandcoolingsection}
The main heating processes inside dense molecular clouds are the
heating  by cosmic rays and heating from shocks. The cosmic ray heating rate
 is \citep{GL78}
\begin{eqnarray}
G_{cr}=6.4\times10^{-28}n(\mathrm{H_2})~\mathrm{ergs~cm^{-3}~s^{-1}}.
\end{eqnarray}
Elastic scattering between two species at different temperatures  will lead to
heat exchange, as well as heat generation if the species are streaming relative
to each other. The heating rate per unit volume produced by elastic scattering
of a species $\alpha$ with a species $\beta$, is \citep{D86}
\begin{eqnarray}\label{heatingrate}
G^{\alpha\beta}=\frac{\rho_\alpha \rho_\beta}{(m_\alpha + m_\beta)} \gamma_{\alpha\beta}[3
\mathrm{k}(T_\beta - T_\alpha) + m_\beta |{\bf v_\alpha -v_\beta}|^2],
\end{eqnarray}
where
\begin{eqnarray}
  \gamma_{\alpha\beta} = \frac{<\sigma v>_{\alpha\beta}}{(m_\alpha +m_\beta)}.
\end{eqnarray}
If a species $j$ is unable to cool efficiently and has a small heat
capacity per unit volume then $G^{jn} =0$ \citep{C87} and
\begin{eqnarray}\label{chargedspeciestemp}
T_j=T+\frac{1}{3\mathrm{k}}m_n|{\bf v}_j-{\bf v}|^2.
\end{eqnarray}
The heating rate of the neutrals due to species $j$  is then
\begin{eqnarray}
G^{nj}=\rho\rho_j \gamma_j|{\bf v}-{\bf v}_j|^2.
\end{eqnarray}
Thus, by equation (\ref{balanceforceequation}), the total heating
rate is
\begin{eqnarray}\label{heatingratefinal}
G^n = \rho\sum_j\rho_j\gamma_j|{\bf v}-{\bf v}_j|^2  = {\bf J \cdot
E'},
\end{eqnarray}
 The low mass of the electrons ensures that their
collisional heating is negligible. With low fractional ionisation
the grains dominate the collisional heating \citep{DRD83}. The total
neutral heating rate is then $G=G^n+G_{cr}$. The dominant cooling
mechanisms inside shock-heated regions is collisional excitation  of
molecular $\mathrm{H_2}$.  Neglecting the vibrational cooling, which
is negligible for $n(\mathrm{H_2})\lapprox10^7\mathrm{cm^{-3}}$
$T\lapprox 3000\mathrm{K}$, we adopt the cooling rate
\begin{eqnarray}\label{leppandshull}
\Lambda = n(\mathrm{H_2}) \frac{L_{rH} L_{rL}}{L_{rH}+L_{rL}} ~ \mathrm{ergs
~cm^{-3}~s^{-1}},
\end{eqnarray}
with the  rotational cooling rate coefficients for  high and low
density, $L_{rH}$ and $L_{rL}$,  from equations (10) and (11) of
\citet{LS83}.

\section{Charged Species}\label{chargedspeciessection}
\subsection{Ions and Electrons}\label{ionsandelectronsection}
The ions  are  singly charged with $m_i=30m_\mathrm{H}$. As
discussed in \citet{W98},  the scattering cross-section $\sigma_i$
has a $1/v$ dependence for drift speeds below $20\mathrm{km/s}$, and
the rate coefficient for ion-neutral scattering is  $<\sigma v>_i
\approx 1.6 \times 10 ^{-9} $cm$ ^3 $s$^{-1}$.  The ion Hall
parameter (equation (\ref{Hallparm})) can thus be written in terms
of the pre-shock Hall parameter
\begin{eqnarray}
\beta_i =\beta_{i0}\frac{B}{B_0}\frac{v_z}{v_s}. \label{ionHall}
\end{eqnarray}
Inside a  molecular cloud \citep{E79}
\begin{eqnarray}\label{fractionalion}
\frac{n_i}{n_H}\approx 10^{-8}\left(\frac{n_H}{10^6\mathrm{cm^{-3}} }
\right)^{-1/2}.
\end{eqnarray}
The ions are unable to cool efficiently, as they have a small heat
capacity per unit volume, due to the low fractional ionisation, thus
$G^{ni}\approx 0$  and equation (\ref{chargedspeciestemp})
applies to $T_i$.
The electron scattering cross-section  is $\sigma_e\approx 1
\times10^{-15}$cm$^{2}$, and as $m_e<<m_n$, then by  equation
(\ref{effvel}):
\begin{eqnarray}
u_e = \left[ \varphi_e + |{\bf v}_e-{\bf v}|^2
\right]^{\frac{1}{2}},
\end{eqnarray}
where
\begin{eqnarray}
\varphi_e=\frac{128}{9\pi}\frac{k T_e}{m_e}.
\end{eqnarray}
The electron temperature $T_e$ is needed to calculate $u_e$ and
$Z_g$. Electrons  and ions are highly coupled to ${\bf B}$
with $|\beta_{i,e}|>>1$ (equation (\ref{Hallparm})), and
thus $({\bf v}_{i}- {\bf v})\approx ({\bf v}_{e}- {\bf v})\approx {\bf E'}\times{\bf B}/B^2 $
(equation (\ref{driftvel})).
Equation (\ref{chargedspeciestemp}), therefore, implies that
$T_e\approx T_i$, however \citet{DRD83} showed that inside shocks in
dense molecular gas the electrons lose energy due to impact
excitation of $\mathrm{H}_2$ and $T_e<T_i$.
$T_e$ is therefore  approximated via (see Fig. 1
of \citet{DRD83})
\begin{eqnarray}\label{teapprox}
T_e \approx T+0.2(T_i - T).
\end{eqnarray}
Once $T_e$, and subsequently  $u_e$, are known,   then $\beta_e$ can be determined.
The calculation of $\beta_e$ is complicated by the dependence of $u_e$ on
 $|{\bf v}_e - {\bf v}|$.  To determine $\beta_e$,
equations (\ref{effvel}), (\ref{driftvel}), and (\ref{Hallparm})
must be solved, this is discussed  in Section
\ref{sectionHallparameters}.

\subsection{Grains}\label{grainsection}
The charge on a dust grain $Z_g$  results from collisions with other charged
particles as well as the ejection of photoelectrons. The charge state of a
grain contains stochastic fluctuations about some mean \citep{S78,
GS75}. If the fluctuations  are neglected, the instantaneous
mean charge on a dust grain of radius $a_g$ is   \citep{D80}
\begin{eqnarray}
Z_g\approx \frac{-4 k T_e a_g}{e^2}. \label{graincharge}
\end{eqnarray}
Equation (\ref{graincharge}) is only valid for
$|Z_g|\gg1$ which is not always satisfied for the pre-shock
conditions of interest; in cold molecular clouds
$T_{e0}=10\mathrm{K}$ and for a MRN distribution $a_g=[50$\AA,
2500\AA],  $|Z_{g0}|<1$. We force  $Z_g =-1$
when $|Z_g|<1$ by equation (\ref{graincharge}). The
scattering cross-section for a spherical grain is $\sigma_g=\pi a^2$,
and by equation (\ref{effvel}),
\begin{eqnarray}\label{graineffvel}
u_g = \left[ \frac{128}{9\pi}\left(\frac{k  T}{m_N} \right)  + |{\bf
v}_g-{\bf v}|^2\right]^{\frac{1}{2}},
\end{eqnarray}
as the grain thermal velocity is negligible compared with that of
the neutrals. The internal density  is
$2.5\,\mathrm{g/cm}^3$.
By equation (\ref{graincharge})  the
relative gyrofrequency ($Z_geB/m_gc$) for a smaller grain
$a_{gs}$ is greater than that of a larger grain $a_{gl}$
by  $(a_{gl}/a_{gs})^2$, and so  the smaller  grains
remain better coupled to the field.
Consequently, the smallest grains have the largest $\beta_g$. The
calculation of $\beta_g$ is addressed  in Section
\ref{sectionHallparameters}.


The abundance of the  pre-shock grains is given by a MRN grain size
distribution \citep{MRN77} which assumes a mixture of silicate and
graphite;
 \begin{eqnarray}
 \frac{dn_{g0}}{da}=A n_H a^{-3.5}, ~~~~~~~A=1.5 \times 10^{-25} \mathrm{cm}^{2.5}. \label{MRN1}
 \end{eqnarray}
The limits are  $[a_1, a_2]=[50\mathrm{\AA}, 2500\mathrm{\AA}]$ and defining $ y = \log a$;
 \begin{eqnarray}
 x_{G0}=\frac{n_{g0}}{n_H} = \int_{a1}^{a2} Aa^{-3.5}da  = \int_{\log{a_1}}^{\log{a_2}}{A}a^{-2.5} dy.\label{MRN2}
 \end{eqnarray}

Inside dense molecular clouds, the grain size distribution may
differ considerably from the MRN model  (e.g. \citealt{LG03}).
Observational measurements of the visible extinction of objects
behind dense clouds suggest that larger dust grains may be present
(e.g. \citealt{WBLBE83}). Inside dense clouds, the growth of ice
mantles and dust coagulation can remove the smaller grains
($a_g<<1000$\AA), and can increase the grain mass by a factor of up
to 2 \citep{CTH93}.  We also consider a MRN distribution with
increased limits $[a_{1m},
a_{2m}]=[90\mathrm{\AA~},4500\mathrm{\AA}]$ \citep{NNU91}. The
distribution is re-normalised with $A_{mantles} a_m^{-2.5} = A
a^{-2.5}$ so $A_{mantles} = A (a/a_m)^{-2.5}=6.5 \times
10^{-25}\mathrm{cm}^{2.5}$ and the mean interior density is
$1.14\,\mathrm{g/cm}^3$.

Using Gauss-Legendre weights (a highly accurate method of integrating smooth functions) the MRN grain size
 distribution may be represented by the summation over  a number of discrete grain
 bins or grain size classes \citep{P96} and  equation (\ref{MRN2}) may be expressed as
\begin{eqnarray}
x_{G0}=\int_{\log{a_2}}^{\log{a_1}} f(y)dy= \sum_{m=1}^N w_m f(y_m)
=\sum_{m=1}^N x_{g0m}, \label{MRN}
\end{eqnarray}
where $y_m$ are abscissae with weights $w_m$, $f(y) = A a^{-2.5}$,
and $ y = \log a$.  The $y_m$ and $w_m$ are pre-determined from the
Legendre polynomial of order N. Each of the N grain bins, have an
associated size $a$, number density $x_{g0m}$, and corresponding
parameters, $Z_{gm}$  $\beta_{gm}$, $v_{gxm}$, $v_{gym}$, and
$v_{gzm}$. Using  equation (\ref{MRN}), any integrals over the grain
size distribution   can be generalised  such that for any function
$g(y)$;
\begin{eqnarray}
\int_{\log{a_2}}^{\log{a_1}}f(y)g(y)dy=\sum_{m=1}^N w_m f(y_m)
g(y_m) =\sum_{m=1}^N x_{gm} g(y_m).
\end{eqnarray}
Equations (\ref{heatingratefinal}) and (\ref{currentdens}) are
calculated in this way, for example;
\begin{eqnarray}
{\bf J} =en_H\left[x_{i}{\bf v}_i -x_{e}{\bf v}_e + \sum_{m=1}^N
x_{g0m}\frac{v_s}{v_{zgm}}Z_{gm}{\bf v}_{gm}\right].
\end{eqnarray}

\section{Hall Parameters}\label{sectionHallparameters}
To determine  $\beta_e$ and $\beta_g$,  equations (\ref{effvel}),
(\ref{driftvel}), and (\ref{Hallparm})  are solved as follows.
Writing
\begin{eqnarray}
{\bf E}'_\parallel = \frac{({\bf B}\cdot {\bf E'})}{B^3}{\bf B}
\end{eqnarray}
and
\begin{eqnarray}
 {\bf E}'_\perp = \frac{{\bf
B} \times({\bf E'}\times {\bf B})}{B^3} = \frac{{\bf
E}'}{B}-\frac{({\bf B}\cdot {\bf E'})}{B^3}{\bf B},
\end{eqnarray}
then  the  drift velocity equation (\ref{driftvel}) may be written as
\begin{eqnarray}
\frac{{\bf v}_j -{\bf v}}{c} = \beta_j{\bf E}'_\parallel+\frac{\beta_j^2}{1+\beta_j^2}\frac{{\bf
E}'_\perp\times {\bf B}}{B^2} +\frac{\beta_j}{1+\beta_j^2}{\bf E}'_\perp. \label{drift2}
\end{eqnarray}
Taking the dot product of ${\bf v}_j -{\bf v}$ with itself gives
\begin{eqnarray}
|{\bf v}_j -{\bf v}|^2=\beta_j^2{{\bf E}^{'2}_\parallel}c^2+\frac{\beta_j^2}{1+\beta_j^2}{\bf
E}^{'2}_\perp c^2. \label{driftHall}
\end{eqnarray}
Defining  $\alpha_j$ as
\begin{eqnarray}
\alpha_j = \frac{Z_jeB}{m_jc}\frac{m_j+m}{\sigma_j\rho}, \label{const2Hall}
\end{eqnarray}
then equation (\ref{Hallparm}) becomes
\begin{eqnarray} \beta_j =
\frac{Z_jeB}{m_jc}\frac{(m_j+m)}{\sigma_ju_j\rho}=\frac{\alpha_j}{u_j},
\label{Hallarange}
\end{eqnarray}
Eliminating $u_j$
from equation (\ref{Hallarange})  using (\ref{effvel}) gives
\begin{eqnarray}
|{\bf v}_j -{\bf v}|^2= \frac{\alpha_j^2}{\beta_j^2} - \varphi_j. \label{Hallsolve}
\end{eqnarray}
After eliminating $|{\bf v}_j -{\bf v}|$ from equations
(\ref{driftHall}) and (\ref{Hallsolve}) then
\begin{eqnarray}
\beta_j^6{\bf E}^{'2}_\parallel &+& \beta_j^4\left({\bf
E}^{'2}_\parallel+ {\bf E}^{'2}_\perp+\frac{\varphi_j}{c^2}\right)
\nonumber \\
 &+& \beta_j^2\left(\frac{\varphi_j}{c^2}-\frac{\alpha_j^2}{c^2}\right)-\frac{\alpha_j^2}{c^2}=0,
\label{Hallsolvecubic}
\end{eqnarray}
which has only  one positive root for $\beta_j^2$. The  root $-|\beta_j|$,
 is then taken for the negatively charged electrons and grains.

\section{Calculating  $E_z'$}\label{EZsection}
Lastly, $E_z'$  is needed, and using equations (\ref{max6})
and (\ref{mass2}),
equation (\ref{driftvel}) is  recast in terms of $v_{jz}$
\citep{W98}:
\begin{eqnarray}
v_{jz}=n_{j0}Z_j(p_jE_z'+q_j), \label{driftagain}
\end{eqnarray}
where
\begin{eqnarray}
p_j&=&\frac{c}{n_{j0}Z_j} \frac{\beta_j}{1+\beta_j^2}
\frac{(B_z^2\beta_j^2+B^2)}{B^3} \label{constp}
\end{eqnarray}
and
\begin{eqnarray}
q_j&=&\frac{1}{n_{j0}Z_j}\left[ v_z +
\frac{c\beta_j^2}{1+\beta_j^2}\frac{({\bf E}'\times{\bf B})_z}{B^2}
\right]  \label{constq} \nonumber \\
  &+&\frac{1}{n_{j0}Z_j}\frac{c\beta_j^3}{1+\beta_j^2}(B_xE_x'+B_yE_y')\frac{B_z}{B^3} .
\end{eqnarray}
From  equations (\ref{max6}), (\ref{mass2}), and (\ref{driftagain}),
\begin{eqnarray}
n_e = \sum_k Z_k n_{0k}\frac{v_s}{v_{jz}}= \sum_k
\frac{v_s}{p_kE_z'+q_k}\label{neequation}
\end{eqnarray}
where $k$ runs over the grain and ion species. Also by equations
(\ref{current3}) and (\ref{currentdens});
\begin{eqnarray}\label{electroneqn}
0=-n_e v_{ez} +\sum_k Z_k n_{k} v_{zk}=-n_e v_{ez} + v_s \sum_k Z_k n_{k0}.
\end{eqnarray}
Using equation (\ref{driftagain}) for  the electrons with  equation  (\ref{electroneqn}), and
using  equation (\ref{neequation})  to eliminate $n_e$ gives
\begin{eqnarray}\label{finalezeqn}
\frac{1}{( p_e E_z' + q_e)}\sum_k \frac{n_{k0}Z_k}{n_{e0}}+\sum_k \frac{1}{p_k E_z' + q_k} =0.
\end{eqnarray}
Equation (\ref{finalezeqn}) has $N$  poles at $E_z'=-q_j/p_j$, with
$N-1$ solutions for $E_z'$. In \citet{W98}, $N=3$ and  equation
(\ref{finalezeqn}) is  an easily solvable quadratic in  $E_z'$. For
arbitrary $N$  the choice of the correct  root for $E_z'$ is not so
simple. Imposing  the requirement $E_z' = 0$ upstream and downstream
enables the identification of the bracketing pole species
($-q_{j0}/p_{j0}$). Since $E_z'$ is  continuous through the shock,
the correct $E_z'$ root will remain between these bracketing species
poles. Since
\begin{eqnarray}
p_{j0}=\frac{c}{n_{j0}Z_{j0}}
\frac{\beta_{j0}}{1+\beta_{j0}^2}\frac{1}{B_0} \left[
\frac{B_z^2\beta_{j0}^2}{B_0^2} +1 \right],
\end{eqnarray}
\begin{eqnarray}
q_{j0}=\frac{1}{n_{j0}Z_{j0}}v_s,
\end{eqnarray}
then $p_{j0}<0$ and the sign of $q_{j0}$  is determined by the sign
of $Z_{j0}$. The ion pole $-q_{i0}/p_{i0}$ will always be negative,
however poles  of the negatively charged species  are positive, thus
$-q_{i0}/p_{i0}$ will always be the minimum bracketing pole.
$\beta_{e0}\sim10^3$ and  $\beta_{g0}\sim 10^1 -10^2$ for the grain
sizes considered, so $|q_{e0}/p_{e0}|< |q_{g0}/p_{g0}|$. Thus the
solution for $E_z'$ is the root of equation (\ref{finalezeqn}) that
lies between $-q_{i}/p_{i}$ and $-q_{e}/p_{e}$.


\section{Shock Calculations}\label{shockcalculationsection}

All fluid variables are made dimensionless by expressing the
velocity, magnetic field, electric field, and  pressure  in units of
$v_s$, $B_0$, $v_sB_0/c$, and $P_0/(\rho v_s^2)$, respectively. The
characteristic length scale of the shock is \citep{W98}
\begin{eqnarray}
 L_s=\frac{v_{A0}}{\sum_j \gamma_{j0}\rho_{j0}}.
\label{lengthscale}
\end{eqnarray}
The shock models are calculated as follows. For the MRN grain size distribution
models,  $x_{g0}$ is  assigned to each grain class (Section \ref{grainsection}).
$x_{e0}$ is then calculated from charge neutrality (equation (\ref{max6})), and $\beta_{e0}$, $\beta_{i0}$ and
 $\beta_{gm0}$ are calculated (equation (\ref{Hallparm})). The appropriate root of the
isothermal jump condition (equation (\ref{isojump}))  for downstream  $v_{zd}$  is
found,  allowing the  calculation of  downstream  $v_x$  $P$, and $B_x$ from equations (\ref{mom1}), (\ref{mom3}) and
(\ref{isocond}). The pole bracketing species for $E_z'$ is
identified as described in Section \ref{EZsection}.

As discussed in \citet{W98},   a one parameter family of solutions
exist and integration from upstream to downstream yields the
intermediate shock solutions, and from downstream to upstream gives
the desired  fast shock solution. Integration of the  ODE's for
$B_x$, $B_y$, and $P$, equations (\ref{current2}), (\ref{current1}),
and (\ref{odepressure}), respectively,  are then performed after
perturbing these quantities from the downstream state. The equations
are stiff so the integration method of \citet{G71} is used. For each
step, the integrator produces values for $B_x$, $B_y$ and $P$,
enabling the calculation of  $v_x$, $v_y$, $v_z$, $E_x'$ and $E_y'$
directly from equations (\ref{mom1})-(\ref{mom3}) and
(\ref{Electx})-(\ref{Electy}). Finding solutions to equation
(\ref{Hallsolvecubic}) for $\beta_e$ and $\beta_{gk}$ and equation
(\ref{finalezeqn}) for $E_z'$, are complicated  by their mutual
dependence. To find a solution, the value of  $E_z'$ obtained in the
last step of  the integrator is used as a starting value.  The
corresponding Hall parameter (equation (\ref{Hallsolve})) is found
(using the updated values of $B_x, B_y,E_x',$ and $E_y'$), including
the corresponding bracketing poles for equation (\ref{finalezeqn}).
The accuracy of this value for $E_z'$ as a root of  equation
(\ref{finalezeqn}) can then be tested. If  tolerances are not met,
the value of $E_z'$ is amended and the process repeated until
required numerical accuracy is reached. Once  $E_z'$ and
$\beta_{i,e,g}$  are known, ${\bf v}_{i,e,g} -{\bf v}$ are
calculated (equation (\ref{driftvel})), enabling the calculation of
the current density (equation (\ref{currentdens})). The heating and
cooling rates, (equations (\ref{heatingrate})  and
(\ref{leppandshull})), can also be found. The right hand sides of
equations (\ref{odepressure}), (\ref{current2}), and
(\ref{current1}), can then be evaluated and passed to the
integrator.

The calculation of radiative shocks is complicated by the long length scales
required over which the shock cools to its post-shock temperature. The
jump conditions yield the downstream state of the shock  but integration cannot
proceed directly from this state as the length scales for the shock to cool
after the other dynamic variables relax is too large.
Relaxation methods have been implemented in the past for this reason
\citep{D80}, requiring  the determination of the fluid variables on grid
points in $B_x-B_y-P$ space and then the variables are `relaxed' until the
correct solution is reached that satisfies the boundary jump conditions. This
method is inapplicable  here since the calculation of  $E_z'$ requires an
initial guess at each point on the shock and an arbitrary choice of $E_z'$ will not always
satisfy bracketing pole criteria of equation (\ref{finalezeqn}).
Integration through the shock however, provides the initial guess for $E_z'$,
which is sufficiently
close to the desired solution and satisfies the pole criteria.

The  alternative we have implemented is to search for the solution
from upstream of the final post-shock state, before radiative
cooling is dominant. An initial starting state for $(B_x^*,
B_y^*,{P_d}^*)$  is chosen in which  $B_x$ and $B_y$ should have
almost obtained their post-shock values; $B_y\approx 0$ and
$B_x^*=B_{xd}+\delta B_x$, where $\delta B_x$ is small ( $\sim 1\%$
of $B_{xd}$). The pressure $P^*=P_{d}+\delta P_d$ is given a larger
relative perturbation and the radiative cooling zone is stepped
over. It is unlikely that integration from this approximate state
will yield the correct solution, but it acts as a good initial
guess.  The calculation is a  two point boundary problem with a
known  desired final (upstream) state   and  the (downstream)
initial values need to be adjusted until integration from downstream
to upstream  yields the correct fast shock solution. This is done by
using the shooting method \citep{P96} which repeatedly performs the
integration for adjusted initial states. The initial starting
parameters ($B_{x^*}$,  $B_{y^*}$, $P_d^*$) are free variables, and
the upstream variables ($B_{x0}$,  $B_{y0}$, $P_0$) are fixed. The
shooting method proceeds by using a multi-dimensional Newton-Raphson
root finder  which zeros three functions obtained by integrating the
ODEs in $B_x$, $B_y$, and  $P$ over the domain of integration.

A requirement of  the shooting method is that for each initial guess
($B_x^*$,  $B_y^*$, $P_d^*$), the subsequent shot can transverse the
entire domain of integration. This is not always the case for very
wrong initial guesses, and physically impossible states may be
reached (e.g., $T$ or $x_e$ may go negative). The convergence of the solution
is complicated by  unsuccessful integration   shots.  Fortunately,
for any number integration shots there was  clear
convergence, before the trajectories diverged.
A multiple shooting method was therefore implemented
which after convergence did occur,  restarts the shooting process at
a point inside the shock using values derived from the converged
solution from the previous shooting procedure. This process is
repeated through the domain of integration, until the upstream state
is reached.

\begin{figure*}
\includegraphics[width=175mm]{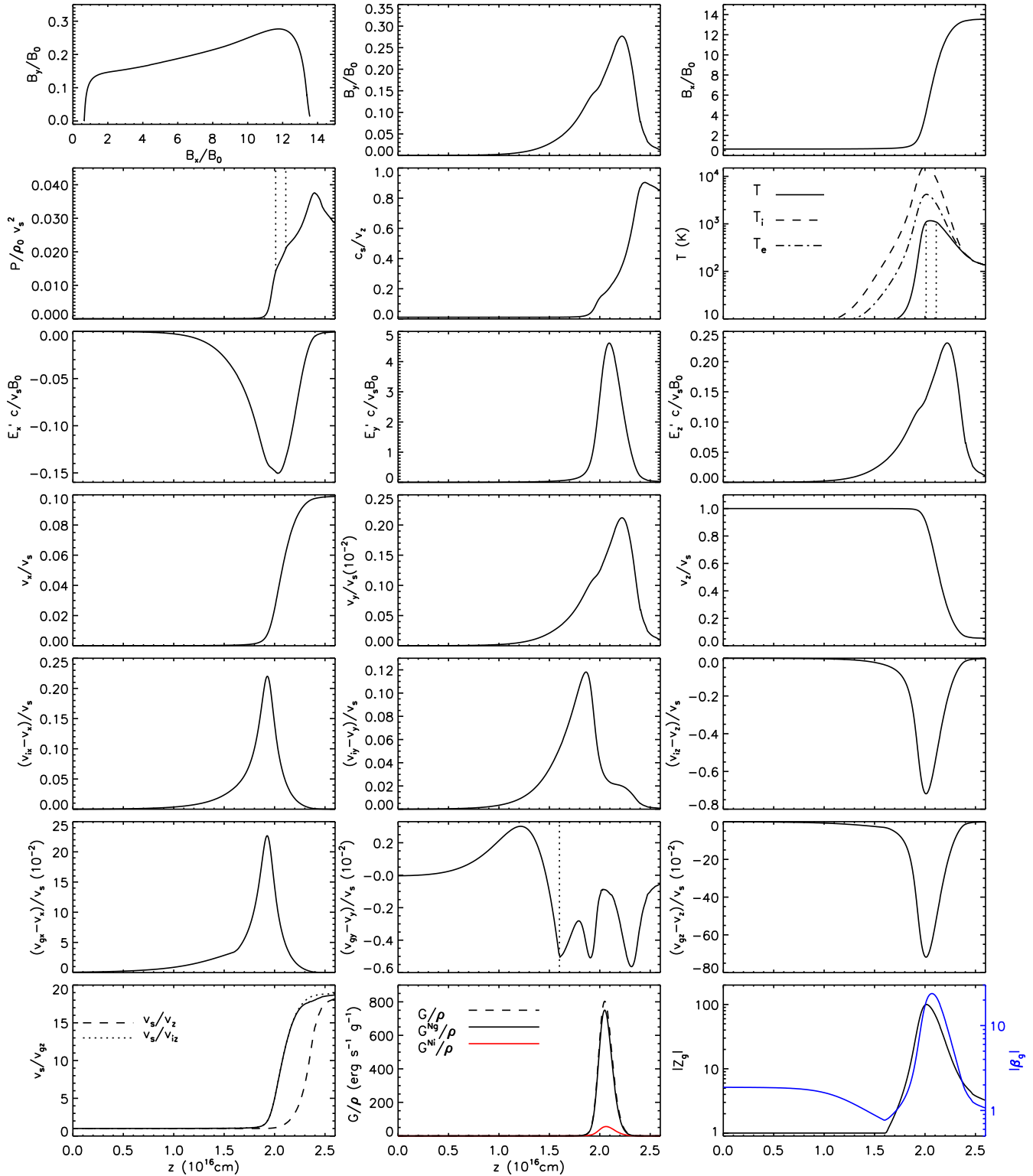}
\vspace{0.5cm}
 \caption{Single sized grain shock model with  grain
radius $0.1\mu \mathrm{m}$, $n_g/n_H =1.6\times 10^{-12}$, and
$\theta=40^\circ$. Shown are the magnetic field components
$B_x,~B_y$, the neutral pressure $P$, ratio $c_s/v_z$ and temperatures $T, T_i$ and $T_e$.
 Also plotted are the electric field components
in the neutral frame $E_x',~ E_y',$ and $E_z'$,  the neutral
velocity components $v_{x,y,z}$,  and   the charged species drift
velocity components $(v_{ix,y,z}-v_{x,y,z})$ and
$(v_{gx,y,z}-v_{x,y,z})$. The grain parameters $Z_g$ and  $\beta_g$,
are shown along with the compression ratios  $v_s/v_{jz}$ and the
heating rates per unit mass,  $G^{nj}/\rho$.}
\label{singlesizesmalltheta40plot}
\end{figure*}

\section{Results: C-type Shock Models}\label{resultssection}
C-type shock profiles are presented assuming  both  small and large
single sized grains (Sections \ref{singlesizedgrainmodel} and
\ref{singlesizedgrainmodelmantles}) as well as the  MRN grain size
distribution  with and without  mantles and PAHs (Sections
\ref{MRNsection},  \ref{MRNmantles} \ref{MRNwithPAHsection}).
$\theta$ is varied in Sections \ref{singlegrainthetacompare} and
\ref{MRNsection}, and the effects of suppressing $B_y$ are
demonstrated in Section \ref{sectionbysuppressed}. Unless stated
otherwise, the pre-shock conditions are:  $B_0=0.3$mG,
$n_H=10^5$cm$^{-3}$,
$n_0=n(\mathrm{H_2})+n(\mathrm{H})+n(\mathrm{H}_e)\approx0.5n_H+0.1n_H=0.6n_H$,
\(n_{i0}/n_H =3\times 10^{-8}\) (by equation (\ref{fractionalion})),
$\gamma=7/5$ (see the discussion near equation (\ref{intenergy})),
and $T_{0}=T_{e0}=T_{i0}=10\mathrm{K}$. The shock speed is
$v_s=18\mathrm{km/s}$

\subsection{Single sized small $a_g=0.1\mu \mathrm{m}$ grain model}\label{singlesizedgrainmodel}

 A single grain species, $a_g=0.1\mu,
\mathrm{m}$ and internal density of $2.5\mathrm{g/cm}^3$, is now
considered. If the total mass of the grains is 0.01 that of hydrogen
\citep{DRD83}, $n_{g0}/n_H=1.6 \times 10^{-12}$ \citep{W98}.
$n_{i0}/n_H=3\times 10^{-8}$ and  $n_{e0}/n_H=2.99984\times 10^{-8}$
by charge neutrality (equation (\ref{max6})). Then $\beta_{i0}=
1.382 \times 10^{4}$, $\beta_{e0} =-3.358 \times 10^{7} $, and
$\beta_{g0} = -1.632$. Fig. \ref{singlesizesmalltheta40plot}  shows
a  shock profile with  $\theta=40^\circ$. The $B_x-B_y$  phase plot
shows the  shock transition from the upstream  state $
(B_{x0},B_{y0})=(0.7,0)B_0$ to the downstream state
$(B_{xd},B_{yd})=(13.7,0)B_0$. The rotation of ${\bf B}_\perp$ out
of and back into the $x-z$ plane is easily identified.

Initially $B_y$ then $B_x$ increase (from upstream to downstream)
near $z=1 \times 10^{16} \mathrm{cm}$, resulting in an increase of
$B_\perp$. As $B_\perp^2/(2B_0^2 M_{A0}^2) + v_z/v_s + P/(\rho_0
v_s^2)$ is constant (by equation (\ref{mom3})), and changes in
$P/(\rho_0 v_s^2)$ are relatively small in this part of the shock
there is a compensatory decrease in $v_z/v_s$. The
compression of the neutral fluid begins, and  $P$ and
 $T$  begin to increase. $T_e$ and $T_i$ also increase, and the approximation used for $T_e$ (equation
(\ref{teapprox})) leads to a maximum of $\sim 4000~\mathrm{K}$,
consistent with \citet{DRD83}.

The ODE for $P$ is sensitive to the compression of the neutral
fluid, and the balance of $G-\Lambda$  with the magnetic gradient
terms of equation (\ref{odepressure}). The $G-\Lambda$ term
dominates in equation (\ref{odepressure}). Initially $P$ increases
as the neutral fluid is compressed and heated. $T$ decreases
 after  $z=2.05\times 10^{16}\mathrm{cm}$, but
$P$ continues to rise since  there is still compression. $P$ does
not fall until $z=2.4\times 10^{16}\mathrm{cm}$ when the compression
slows, and $v_z/v_s$ reaches its downstream value. $T$ and $P$ have
not yet reached their post-shock states in Fig.
\ref{singlesizesmalltheta40plot}, and will not reach them until far
downstream of the shock, after the neutral fluid cools inside the
radiative zone.

The $P$  profile  shows two  discontinuities in $dP/dz$ at $z=2
\times 10^{16} \mathrm{cm}$ and $z=2.1 \times 10^{16} \mathrm{cm}$,
indicated by dotted lines. These occur at $T=1087 \mathrm{K}$ and
are artefacts of the  cooling function (equation
(\ref{leppandshull})) which is discontinuous  at $T=1087 \mathrm{K}$
\citep{LS83}. Interpolation of  $\Lambda$ was used to smooth the
transition at $T = 1087 \mathrm{K}$, however $P$  is highly
sensitive to $\Lambda$, still reflecting the discontinuity in
$\Lambda$.

Fig. \ref{singlesizesmalltheta40plot}  shows  the ratio $c_s/v_z$
nearing   the sonic condition  at $z =2.45 \times 10^{16}
\mathrm{cm}$ ($c_s/v_z\sim0.9$). The peak coincides with the slowing
of the compression as $v_z$ approaches the downstream value.
$c_s/v_z$ then decreases as $c_s$ decreases with $T$.   $v_z=c_s$ if
the neutral fluid cools quickly enough after the peak in $T$. This
occurs for certain pre-shock conditions, e.g., high pre-shock number
densities and/or low $\theta$. In this case, the shock becomes
either $\mathrm{C}^*$-type or J-type, requiring different numerical
procedures  to calculate the shock structure since equation
(\ref{odepressure}) cannot be integrated through the critical point
(\citet{RD90}).

In Fig.\ref{singlesizesmalltheta40plot}, $E_x'$ and $E_z'$ are
similar in magnitude, with $E_y'$  an order of magnitude larger.
Since   $v_x \propto (B_x-B_{0x})$ and $v_y\propto B_y$ (equation
(\ref{mom2})), trends in  $v_x$ and $v_y$ follow those of $B_x$ and
$B_y$. $v_x$ steadily increases through the shock, before reaching
the downstream state given by the jump conditions.  The magnitude of
$v_y$ is orders of magnitude smaller than $v_x$ and $v_z$. As  ${\bf
E'}_\parallel =({\bf B}\cdot{\bf E'}){\bf B}/B^3$ is negligible
through the shock, and by equation (\ref{driftvel})
\begin{eqnarray} \label{driftapprox}
\frac{{\bf v}_{j}-{\bf v}}{c} &\approx&
\frac{\beta_j^2}{1+\beta_j^2}\frac{({\bf E}'\times {\bf B})}{B^2} +
\frac{\beta_j}{1+\beta_j^2} \frac{{\bf E'}}{B}.
\end{eqnarray}
For $|\beta_j|>>1$  the species are tied to ${\bf B}$
with $ ({\bf v}_{j}-{\bf v})\approx ({\bf E}'\times
{\bf B})c/B^2$, as is the case for the ions and electrons. The ions
(and electrons) lead the neutrals in the $x$ and $y$ directions but
lag the  neutrals in $z$. The second term in equation
(\ref{driftapprox}) becomes non-negligible for the grains with
$|\beta_g|\sim1-10$. $(v_{iy}-v_y)$ has more structure than both
$(v_{ix,z}-v_{x,z})$. The shoulder at $z\sim2.2\times
10^{16}\mathrm{cm}$ where $(v_{iy}-v_y)$ declines much less
rapidly, corresponds to the peak of $E_z'$  and a slowing down of
the compression of $B_x$ ($B_x\approx B_{xd}$). As $E_z'$
subsequently decreases after $z=2.25 \times 10^{16}\mathrm{cm}$,
there is a corresponding reduction  in $(v_{iy}-v_y)$.

\begin{figure*}
\includegraphics[width=175mm]{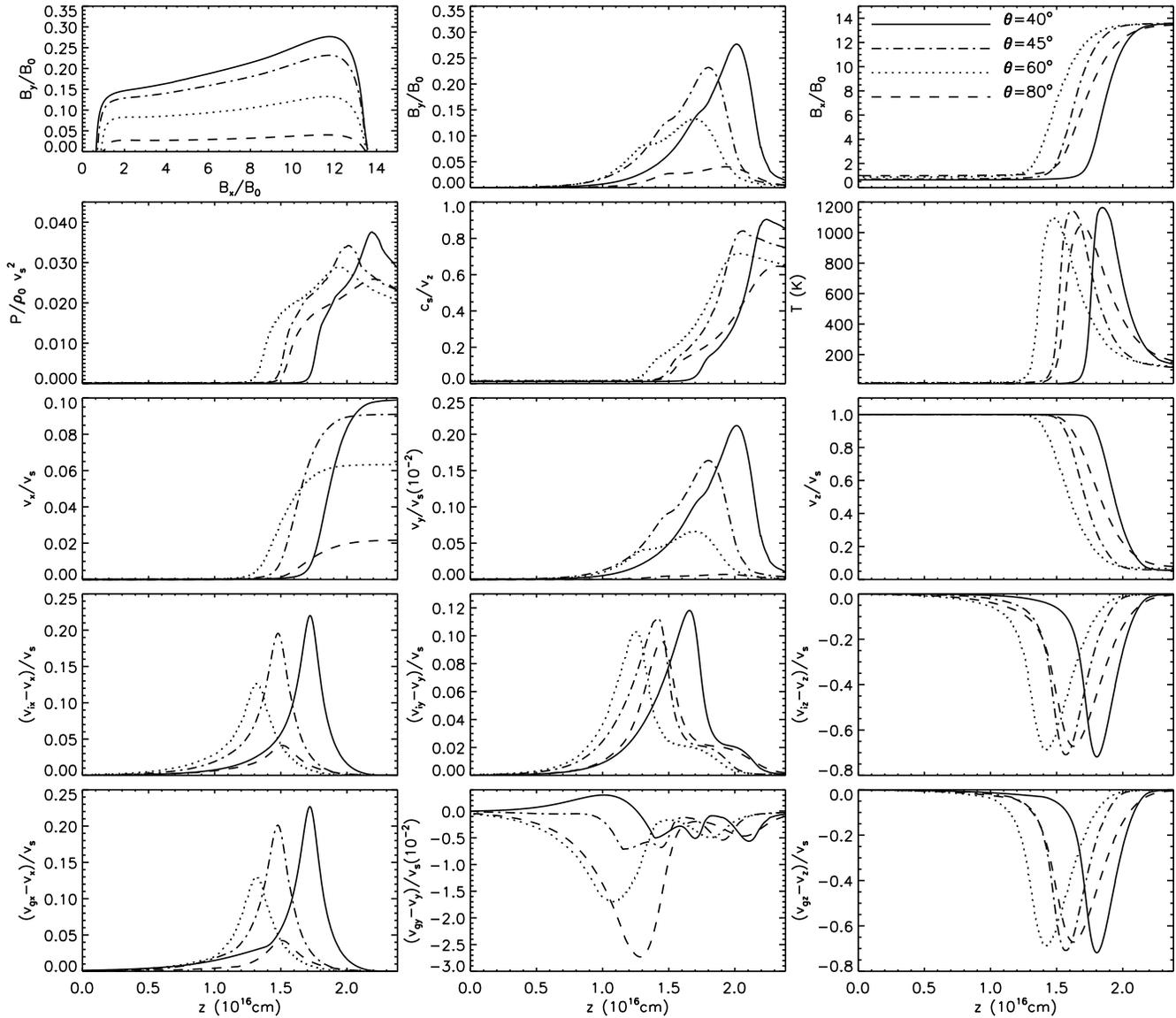}
\vspace{0.5cm} \caption{Comparison of $\theta=40^\circ$ (solid),
$45^\circ$ (dot-dash), $60^\circ$ (dotted), and $80^\circ$ (dashed)
for the $a_g=0.1\mu \mathrm{m}$
and $n_g/n_H =1.6\times 10^{-12}$ model. The $\theta=40^\circ$ line series
is the same as in Fig. \ref{singlesizesmalltheta40plot}.
The descriptions of the quantities are  the same as in the  caption
for Fig. \ref{singlesizesmalltheta40plot}.}
\label{singlesizesmallthetacompareplot}
\end{figure*}

$(v_{gx,z}-v_{x,z})$ are shown in Fig.
\ref{singlesizesmalltheta40plot}, and are  similar in magnitude to
$(v_{ix,z}-v_{x,z})$. $(v_{gy}-v_{y})$ is an  order of magnitude
smaller then $(v_{iy}-v_{y})$ and contains oscillations. The  ratios
$v_s/v_{jz}$ (equations
(\ref{mass1}) and (\ref{mass2})) show that the  grains and ions are
compressed ahead of the neutrals at $z=1.8 \times 10^{16}
\mathrm{cm}$ (along with ${\bf B}$), with compression in
the neutral fluid occurring after $z=2.0 \times 10^{16}\mathrm{cm}$.


 The grain charge remains $-1$ until $T_e$ increases and $|Z_g|>1$ by equation (\ref{graincharge}).
The peak in $|Z_g|$  corresponds to the peak in $T_e$. Initially
 $|\beta_g|\sim 1$ , but  as the grain drift increases,  $u_g$ increases (equation (\ref{effvel})) and
there is a decrease in $|\beta_g|$. $|\beta_g|$ starts to increase
when $|Z_g|$ increases. As ${\bf B}$ is  compressed
$|\beta_g|$ increases further and the grains become better
coupled to ${\bf B}$.  The decrease in $|\beta_g|$ after
$z=2.1 \times 10^{16}\mathrm{cm}$ follows the decrease in $|Z_g|$ as
$T_e$ declines.

As $|\beta_g| \sim 1-10$,   ${\bf v}_g-{\bf v}$ is dependent on the
balance of terms in equation (\ref{driftapprox}). When
$|\beta_g|\approx1$ the second term reaches a maximum for a given
${\bf B}$ and ${\bf E'}$, and ${\bf v}_g-{\bf v}$ is determined by
the balance of both terms. For $|\beta_g|<1$  both terms decrease
with $|\beta_g|$ and $|{\bf v}_g-{\bf v}|$ decreases, signifying an
increase in the grain-neutral coupling. The oscillations in
$(v_{gy}-v_{y})$ are due to the balance of the competitive terms in
equation (\ref{driftapprox}). The sharp increase in
$(v_{gy}-v_{y})$, marked by the dotted line, corresponds to the
increase in  $|Z_g|$ and $|\beta_g|$.  $G^{ng}$ dominates $G$  in
Fig. \ref{singlesizesmalltheta40plot} with only a small contribution
from $G^{ni}$.


\subsection {$a_g=0.1\mu \mathrm{m}$ grain model: varying
$\theta$}\label{singlegrainthetacompare}

A comparison of the $a_g =0.1\mu \mathrm{m}$ models for
$\theta=40^\circ,~ 45^\circ, ~60^\circ,$ and $\theta=80^\circ$, is
plotted in  Fig. \ref{singlesizesmallthetacompareplot}. The $z$
offset in each $\theta$ model is arbitrary, and the solutions have
been shifted in $z$  to overlay and compare. Since
$B_{x0}=B_0\sin\theta$ the upstream state differs  for each model.
As $\theta$
decreases from $80^\circ$ to $40^\circ$, the amount of ${\bf
B}_\perp$ rotation increases as seen in the $B_x-B_y$ phase plot.. For the $\theta=80^\circ$
model, $B_y$ and $v_y$ are small, and  the neutrals are restricted
mainly to the $x-z$ plane. Conversely, for $\theta=40^\circ$ there
is maximum rotation of ${\bf B}_\perp$ with largest $B_y$ and $v_y$.
The shock widths  do not differ greatly with $\theta$.  There are
larger peaks in $P$ and $T$ as $\theta$ is decreased. The neutrals
approach a sonic point as  $\theta$ decreases, with the peak $c_s/v_z \rightarrow 1$.
 $\mathrm{C}$-type solutions are restricted by the
choice of $\theta$,  $\theta_{critical} \sim 40^\circ$, below
which the neutral fluid becomes subsonic inside the shock.

The jump conditions are $\theta$ dependent so the downstream values of
$v_{x}$ and $v_{z}$ differ for each $\theta$ model in Fig.
\ref{singlesizesmallthetacompareplot}.  The differences in
  $B_y$ and $B_z$  with $\theta$  leads to
differences in  $(v_{ix}-v_{x})$, e.g., when $\theta
=80^\circ$, the peaks in  $B_z$ and $B_y$, and thus $({\bf E}'\times
{\bf B})_x$ are small, leading to a lower peak $(v_{ix}-v_x)$  (equation
(\ref{driftapprox})). $(v_{iy,z}-v_{y,z})$ show smaller differences, with
the  largest peak in $|v_{iy,z}-v_{y,z}|$ for the low $\theta
=40^\circ$ case.

$(v_{gx,z}-v_{x,z})$ are similar in magnitude to
$(v_{ix,z}-v_{x,z})$, showing the same $\theta$ dependent features.
Differences in  the structure of $(v_{gy}-v_{y})$ with  $\theta$ are
large, and as $\theta$ increases from $40^\circ$ to $80^\circ$  the
number of oscillations decreases and the relative peak magnitude
increases. $|Z_g|$ and $|\beta_g|$ are comparable for all $\theta$
(see Fig. \ref{singlesizesmalltheta40plot}),  and differences in
$(v_{gx,y,z}-v_{x,y,z})$ are mainly due to the differences in ${\bf
E'}$ and ${\bf B}$.


\subsection{Single sized large $a_g=0.4\mu \mathrm{m}$ grain model}\label{singlesizedgrainmodelmantles}

\begin{figure*}
\includegraphics[width=175mm]{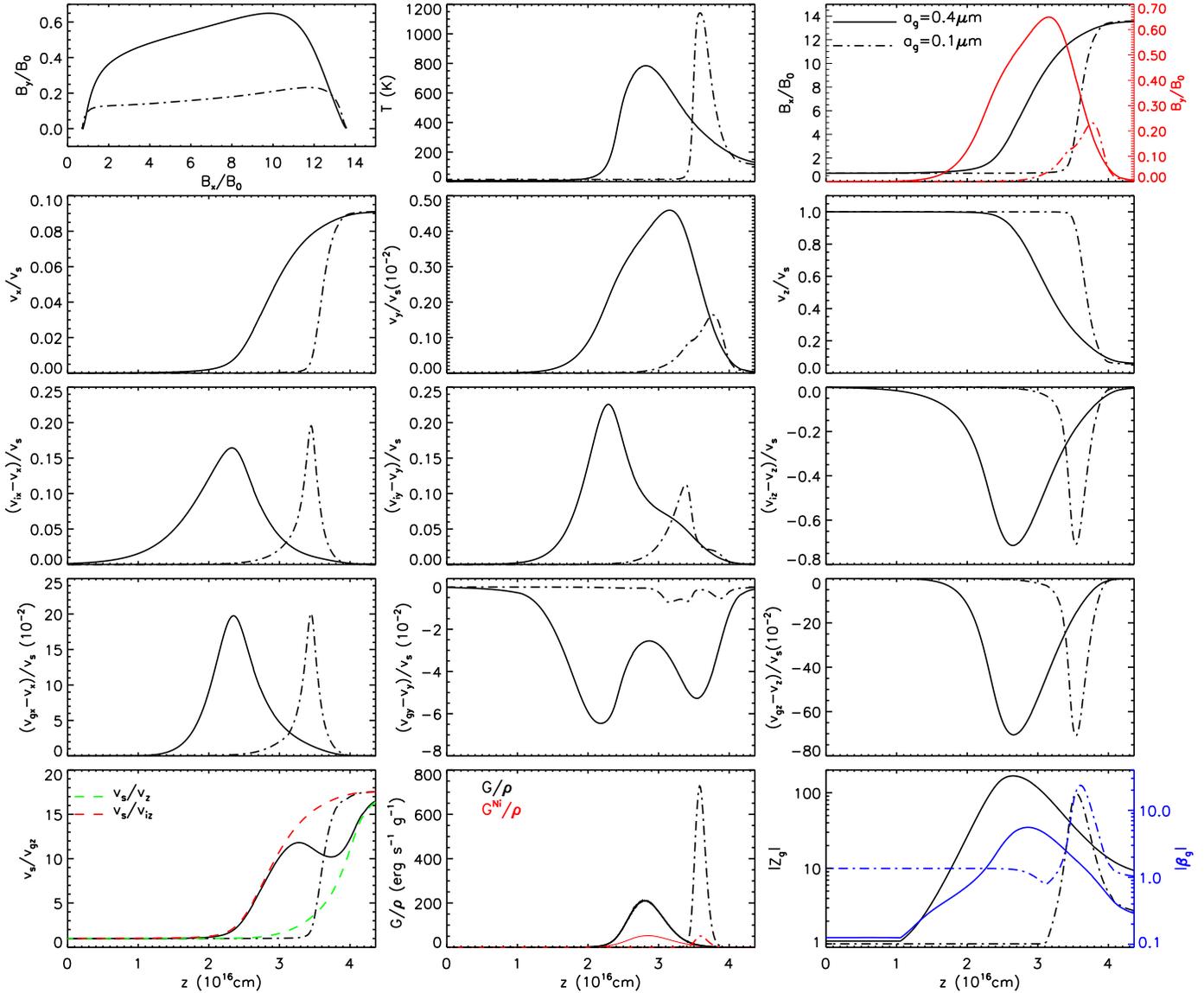}
\vspace{0.5cm} \caption{Comparison of $\theta=45^\circ$ single size
grain models;  $a_g=0.1\mu \mathrm{m}$ with $n_{g0}/n_H =1.6\times
10^{-12}$ (dot-dash)  and $a_g=0.4\mu \mathrm{m}$ with $n_{g0}/n_H
=2.5\times 10^{-14}$ (solid). The descriptions of the corresponding
quantities are the same as in  the caption of Fig.
\ref{singlesizesmalltheta40plot}.} \label{compareSSwithGMsinglesize}
\end{figure*}

Dust grains inside dense molecular clouds are larger than those in
the diffuse interstellar medium (e.g., \citealt{CSS73}) due to
coagulation and the growth of  ice mantles. The grain size is now
increased to $a_g=0.4\mu \mathrm{m}$ to represent grain growth via
coagulation. Assuming the total mass of the grains is
0.01 that of hydrogen then $x_{g0}=2.5\times 10^{-14}$
\citep{DRD83} and  $\beta_{g0}=-0.1020$. $x_{g0}$ and  the net
grain charge is decreased by two orders of magnitude compared
with the $a_g = 0.1\mu \mathrm{m}$ model. The      $a_g=0.1\mu \mathrm{m}$
and       $a_g=0.4\mu \mathrm{m}$  models for $\theta =
45^\circ$ are compared  in Fig. \ref{compareSSwithGMsinglesize}.
The $0.4\mu \mathrm{m}$ grains are less coupled to ${\bf B}$, so the rotation of ${\bf B}_\perp$ increases.
 The shock width is larger and there is a corresponding  reduction in the peak
 temperature to $T \sim 800\mathrm{K}$, as the same amount of kinetic energy flux
must be dissipated into thermal energy as for the thinner
$0.1\mu{\mathrm{m}}$ model shock.

The remaining variables plotted in Fig.
\ref{compareSSwithGMsinglesize} are consistent with the field
components, e.g., there are larger peaks in $|v_{iy,gy}-v_{y}|$ and $v_y$,
consistent with larger $B_y$. There is also  a steeper decrease
(increases) in $v_z$ ($v_x$) for the smaller grain case then for the
$a_g=0.4\mu\mathrm{m}$ case.  The maximum peaks of
$(v_{ix,z}-v_{x,z})$ and  $(v_{gx,z}-v_{x,z})$  are similar for both
models.

\begin{figure}
\includegraphics[width=90mm]{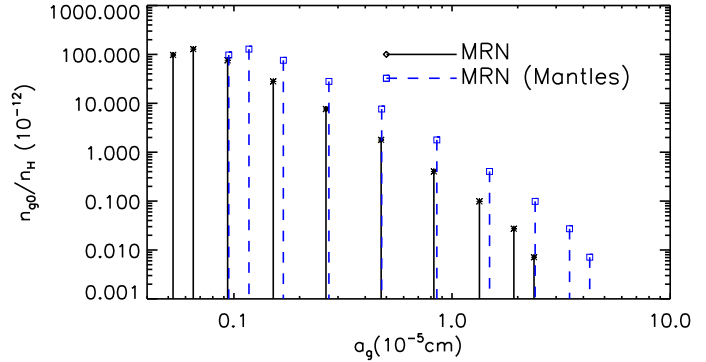}
\vspace{-0.5cm} \caption{Grain radius $a_g$ plotted with
$x_{g0}=n_{g0}/n_H$ for a standard MRN distribution (equation
(\ref{MRN})) and a MRN distribution with grain coagulation and ice
mantles using 10 grain size classes.} \label{MRNgrainsizesplots}
\end{figure}

\begin{figure*}\label{MRNtheta45plot}
\includegraphics[width=175mm]{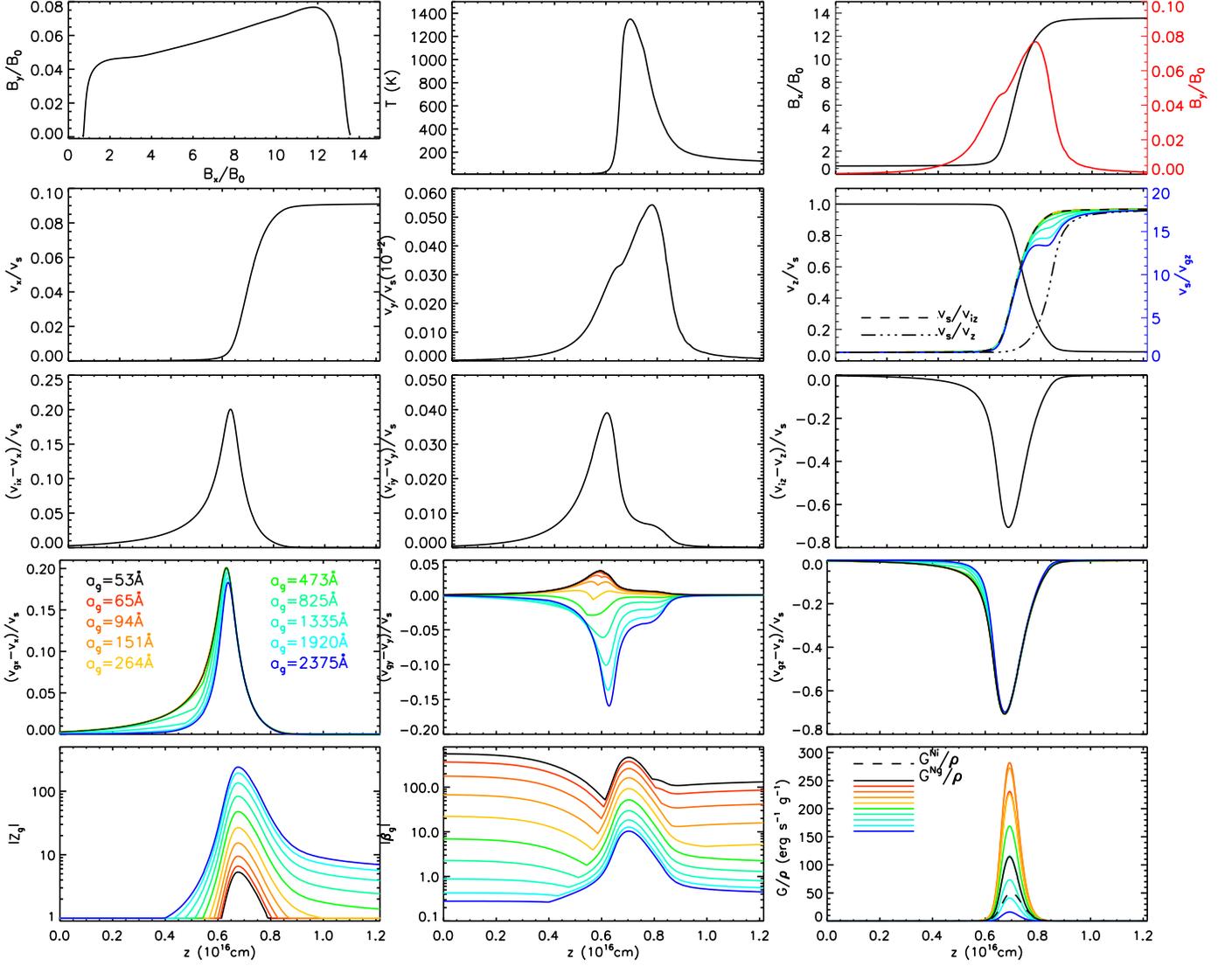}
\vspace{-5.2cm} \caption{Shock profile with  $\theta=45^\circ$ and
the MRN grain size distribution of Fig. \ref{MRNgrainsizesplots}.
The descriptions of the quantities are  the same as for Fig.
\ref{singlesizesmalltheta40plot}. The grain  parameters for each
size class are plotted in a  different colour, with the radius $a_g$
indicated in the $(v_{gz}-v_z)$ plot.} \label{MRNtheta45plot}
\end{figure*}

The decoupling of the $0.4\mu \mathrm{m}$ grains is clearly seen in
the ratio $v_s/v_{gz}$ (equation (\ref{mass2})). Initially the
grains are compressed along with ${\bf B}$  and the ion fluid
($v_s/v_{gz}\sim v_s/v_{iz}$) until after $z\sim3\times 10^{16}
\mathrm{cm}$  when $|\beta_g|$ and $|Z_g|$ decrease. In which case,
the grains are  decoupled from ${\bf B}$  and the compression slows
compared with  the ions. The grain fluid  undergoes
expansion until $z=3.7\times 10^{16}\mathrm{cm}$. Compression then
begins  again, shown by increasing $v_s/v_{gz}$, with the
transition of $|\beta_g|$ from $|\beta_g|>1$ to $|\beta_g|<1$ as the
grains become completely decoupled from ${\bf B}$, and compressed
along with the neutrals after $z=3.7\times 10^{16}\mathrm{cm}$.

A lower grain abundance for the $0.4\mu \mathrm{m}$ model leads to a
reduced  heating rate per unit mass $G^{ng}$, whereas $G^{ni}$ are
similar for both models.  The neutral temperature $T$ (and
$T_{i,e}$), reaches smaller peak values, however the
$0.4\mu\mathrm{m}$ grains obtain larger charge $|Z_g(a_g,T_e)|$  as
they have a larger surface area for electrons to stick to (equation
(\ref{graincharge})).

\subsection{MRN grain size distribution}\label{MRNsection}

In this section a MRN  grain size distribution is  adopted with 10
grain size classes.   $x_{g0}=n_{g0}/n_H$ versus size class $a_g$,
(equation (\ref{MRN}))  is plotted in Fig. \ref{MRNgrainsizesplots}
for the standard MRN grain size distribution. The total pre-shock
grain density is $\sum_g n_{g0} = 3.39\times 10^{-5}
\mathrm{cm^{-3}}$, leading to $x_{e0}=x_{i0}-\sum_g x_{g0}
=2.966\times 10^{-8}$ (equation (\ref{max6})).  In general, the
smaller $a_g$, the larger $x_{g0}$. The only exception is the
$65$\AA~ class with $x_{g0}$ greater than that of the  $53$\AA~
class, since for smaller $a_g$ the Gauss-Legendre weight $w_m$ is
less than the corresponding $w_m$ for midrange $a_g$ (equation
(\ref{MRN})). A significant proportional ($67\%$) of the total grain
number density $\sum_g n_{g0}$ comes from the smaller $a_g\le
65$\AA~ grains, which couple best to the magnetic field and suppress
${\bf B}$ diffusion. As the number of grain bins is increased from 1 to 10,
the shock solution converges; The shock solution is already
converged for $>10$ grain bins (plots not shown here).

Fig. \ref{MRNtheta45plot} shows the corresponding shock profile. The
maximum peak in $B_y$ is  significantly smaller than that of the
equivalent $0.1\mu \mathrm{m}$ grain model (Fig.
\ref{singlesizesmallthetacompareplot}). The larger (low $|\beta_g|$)
size classes are better coupled to the neutrals through collisions,
but their abundances are too low to have any large effect on the
overall structure. The small dominant grain  size classes
(large $|\beta_g|$) remain coupled to ${\bf B}$ and so there is
minimal rotation of ${\bf B}$ out of the $x-z$ plane, and
subsequently a lower peak in $B_y$.  The shock is also thinner and
the peak in $T$ is higher, compared with the single grain models.

\begin{figure*}
\includegraphics[width=175mm]{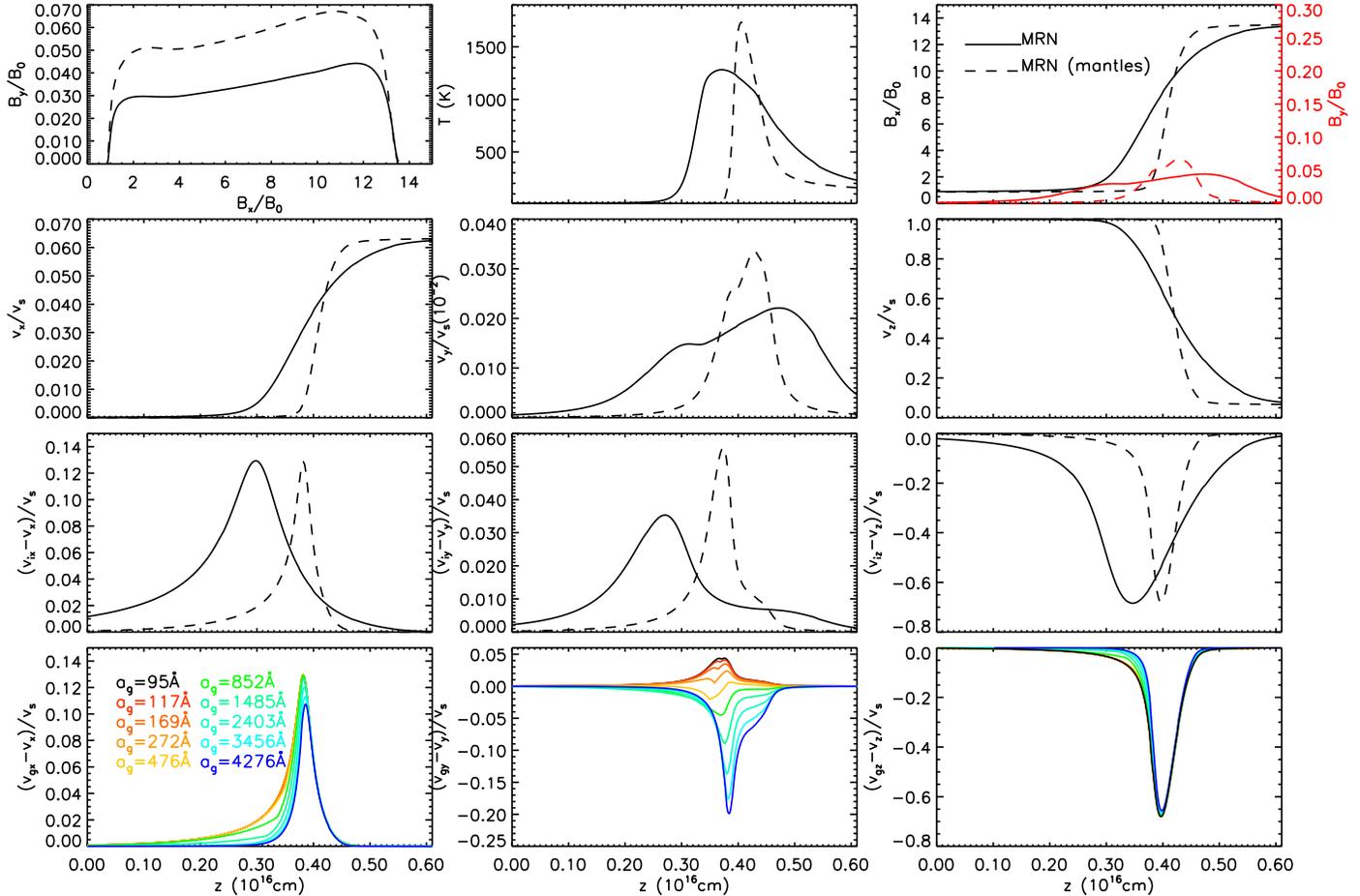}
\vspace{-2.50cm} \caption{ Shock profiles for the  MRN grain size
distribution (solid line) and MRN a distribution with mantles
(dashed line), for $\theta=60^\circ$ and $x_{g0}$ from Fig.
\ref{MRNgrainsizesplots}. $(v_{gx,y,z} - v_{x,y,z})$  are shown for
the MRN(mantles) model only. The descriptions of the quantities are
the same as in the caption for Fig. \ref{MRNtheta45plot}.}
\label{MRNvsMRNgmtheta45}
\end{figure*}

In Fig. \ref{MRNtheta45plot}, $(v_{gx}-v_x)$ clearly differs with
$a_g$, and  the smallest grains are accelerated ahead of the larger
grains until  the peak at $ z= 0.63 \times 10^{16} \mathrm{cm}$. The
 $|\beta_g|$ plot shows for the smallest grains $|\beta_g|>>1$ through the entire shock with
 $(v_{gx}-v_x)\sim(v_{ix}-v_x)$. Conversely, for mid size range grains  with $|\beta_g|<<1$ initially, $|\beta_g|$
reaches a maximum of only $\sim 10$ due to increases in $|Z_g|$. The
largest  grains have the  greatest surface areas, and so obtain
largest $|Z_g|$. As $|\beta_g|$ increases, the larger grains become
better coupled to ${\bf B}$ and $(v_{gx}-v_x)_{large\,
grain}\rightarrow (v_{gx}-v_x)_{small\, grain}$. At $z=0.63 \times
10^{16} \mathrm{cm}$, $(v_{gx}-v_x)$ is a maximum for all sizes,
 at which point $v_x$ increases, and $(v_{gx}-v_x)$ subsequently
decreases, and  all grains classes are (near) uniformly decelerated
in $x$ with respect to the neutrals. Similar comments can be made in
regard to the magnitudes of $(v_{gz} - v_z)$.

$(v_{gy}-v_y)$  differ substantially with $a_g$ in both sign and
magnitude. The smallest $a_g$ ($|\beta_g|>>1$) class have
$(v_{gy}-v_y)\sim(v_{iy}-v_y)$ as expected, and are accelerated
ahead of the neutrals. However the largest $a_g$
($|\beta_g|\sim1-10$) class obtains large negative $(v_{gy}-v_y)$
 (as was also observed for
the single $0.4\mu{\mathrm{m}}$ model). For each of the larger
grains $a_g\geq 825$\AA~ there is a sharp trough in $(v_{gy}-v_y)$
which coincides with $|\beta_g|\sim1$, after which
 as $|\beta_g|$ increases, the grains become better coupled to
${\bf B}$  and $|v_{gy}-v_y|$ decreases.

\begin{figure}
\hspace{1cm}
\includegraphics[width=80mm]{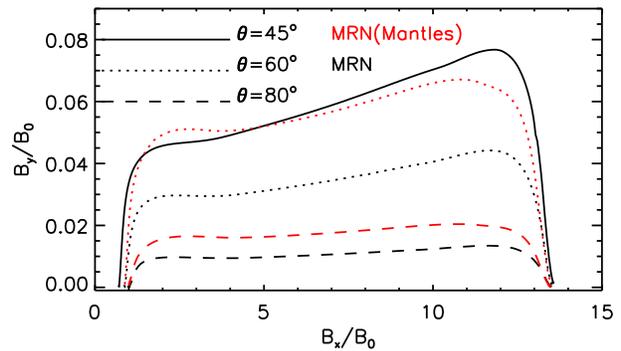}
\vspace{0.5cm} \caption{$B_x-B_y$ phase plot for the  MRN  and
MRN(Mantles) models, plotted in black and red respectively, and
varying $\theta=45^\circ$(solid), $60^\circ$(dotted), and
$80^\circ$(dashed). } \label{MRNthetacomparephaseplot-MRNvsmantles}
\end{figure}

\begin{figure*}
\includegraphics[width=175mm]{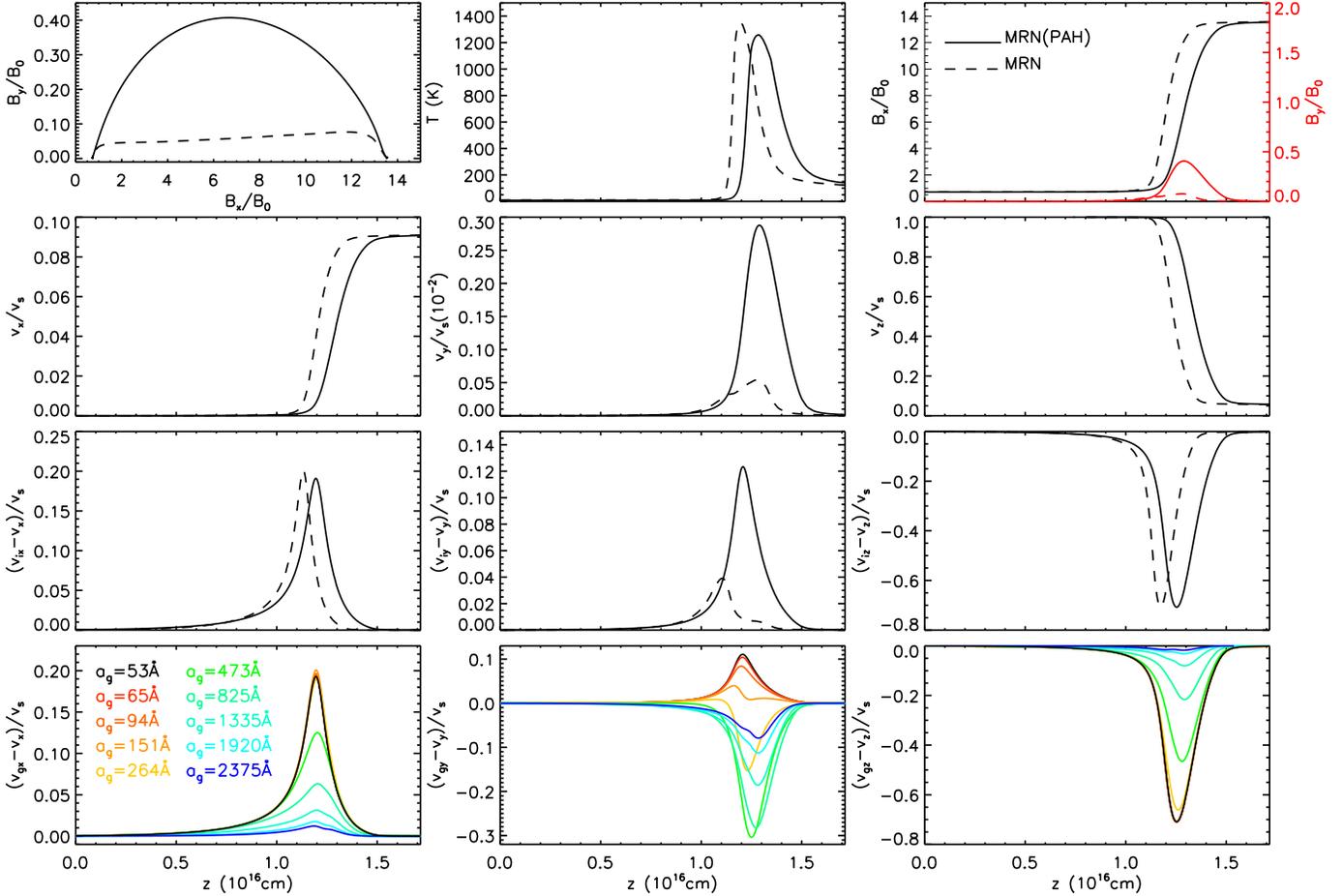}
\vspace{-2.5cm} \caption{Shock profile for the MRN(PAH) model with
$\theta=45^\circ$ (solid line). Also plotted is the MRN model from
Fig. \ref{MRNtheta45plot} (dashed line). $(v_{gx,y,z} - v_{x,y,z})$
and $v_s/v_{zj}$ are shown for the MRN(PAH) model only. The
descriptions of the quantities are  the same as in the caption for
Fig. \ref{MRNtheta45plot}. } \label{mrntheta45withPAH}
\end{figure*}

Each grain class is differentially compressed as shown in the
$v_s/v_{gz}$ plot (equation (\ref{mass2})) in Fig.
\ref{MRNtheta45plot}. The compression of the larger $a_g$ classes
begins to slow ($d(v_z^{-1})/dz\sim0$) near $z=0.8 \times 10^{16}
\mathrm{cm}$, before increasing as $v_{gz}\rightarrow v_{zd}$.
Initially as $|\beta_g|$ increases, the larger grain classes become
better coupled to ${\bf B}$ and are compressed along with the ions
and ${\bf B}$.  The compression then slows with decreasing
$|\beta_g|$ after $z=0.7\times10^{16}\mathrm{cm}$, and the larger
grains decouple from ${\bf B}$. As $\beta_g\approx 1$ the
compression slows ($d(v_z^{-1})/dz\sim0$ at $z=0.8 \times 10^{16}
\mathrm{cm}$), after which $|\beta_g|$ decreases further
 and the large grain classes are compressed
along with the neutrals (see also Fig.
\ref{singlesizedgrainmodelmantles}). In general, the peak in
$G^{Ng}/\rho$, also shown  in Fig. \ref{MRNtheta45plot}, increase
 with decreasing $a_g$  (due to increasing $x_g$), and the smallest grains dominate the heating. $G^{ni}$ is comparable
to that of the $a_g=1920$\AA~ class.

\subsection{MRN grain size distribution with
mantles}\label{MRNmantles}

The MRN distribution  with ice mantles  (discussed near equation
(\ref{MRN2})) is now considered. $x_{g0}$  for each size class are
shown in Fig. \ref{MRNgrainsizesplots}  with $\sum_g n_{g0} =
3.39\times 10^{-5} \mathrm{cm^{-3}}$. A comparison of shock profiles
for $\theta=60^\circ$, using  MRN distributions, with and without
grain mantles are shown in Fig. \ref{MRNvsMRNgmtheta45}. The
rotation of ${\bf B}_\perp$ is  increased for the MRN(mantles) case
as the grains size classes are larger and are only partially coupled
to ${\bf B}$. The shock width is  narrower with a corresponding
higher peak $T$.  A higher peak in $T$ leads to higher $c_s$, and shock
solutions are limited to
 $\theta_{crit}\sim 60^\circ$ for
the MRN(mantles) model, as opposed to   $\theta_{crit}\sim 45^\circ$ for the MRN case.
 A larger peak in $B_y$ leads to  higher
peaks in $v_y$ and $(v_{iy}-v_y)$. $(v_{ix,z}-v_x,z)$ and $v_x,z$
are similar in magnitude for both models. In general the grain drift velocities show similar
trends  to the  MRN model (Fig. \ref{MRNtheta45plot}).

The $B_x-B_y$ phase plots for the  MRN  and
MRN(mantles) models  with varying $\theta$ are compared in Fig.
\ref{MRNthetacomparephaseplot-MRNvsmantles}.  In general, there is a decrease
in the rotation of ${\bf B}_\perp$ with increasing  $\theta$. The
corresponding effects on  the neutral and charged species dynamics
follow on from the relative magnitude of $B_y$ through the shock.
 MRN(mantles) $\theta =60^\circ$ is  essentially  equivalent to MRN $\theta =45^\circ$.
Increasing the grain sizes, which leads to larger forces out of the
shock plane and a more non-coplanar solution, is essentially
equivalent to taking a MRN distribution with smaller $\theta$. This
result is initially surprising as one might expect the addition of
more highly coupled charge carriers with $|\beta_g|>>1$ to reduce
the Hall current. However electron abundance and therefore net
charging of grains is reduced, so many grains remain decoupled from
${\bf B}$.
\begin{figure*}
\includegraphics[width=175mm]{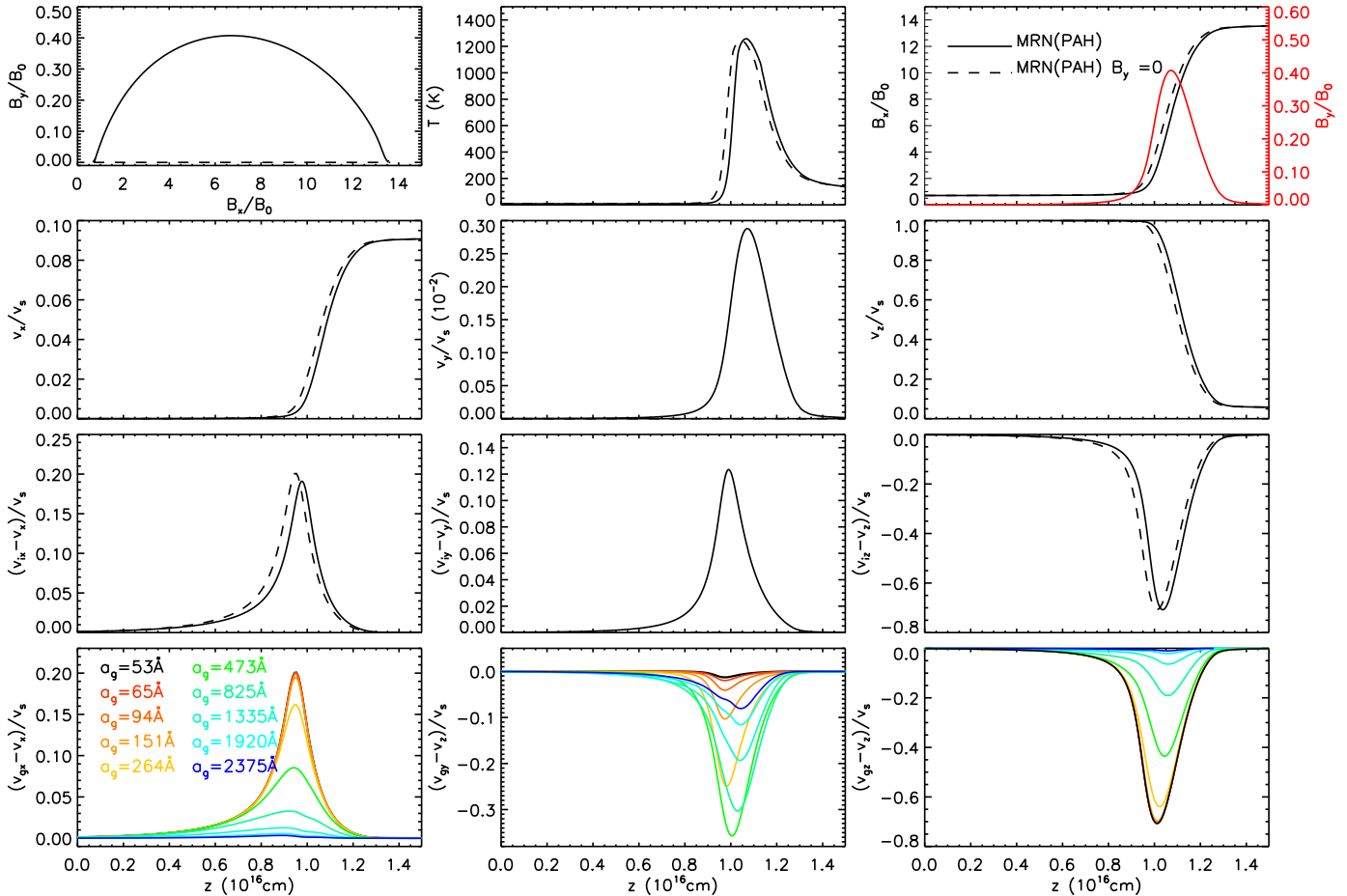}
\vspace{-2.5cm} \caption{ Shock profile for MRN(PAH) model $B_y=0$
(dashed) and $\theta=45^\circ$. The solid line is the MRN(PAH) model
 (Fig. \ref{mrntheta45withPAH}).  $(v_{gx,y,z} -v_{x,y,z})$  are shown
for the MRN(PAH) $B_y=0$   model only.}
\label{mrntheta45withPAHwithbysup}
\end{figure*}

\subsection{MRN with PAHs}\label{MRNwithPAHsection}

In the interstellar medium, for $n_H=10^5\mathrm{cm}^{-3}$, the
abundance of negatively charged PAHs$^-$ is of the order $10^{-8}$,
with a similar total abundance, $1.1\times 10^{-8}$, of ions and
positively charged PAH$^+$s (values taken from Fig. 1 of
\citet{KN96}). Assuming a MRN distribution as before, $\sum_g x_{g0}
= 3.39\times 10^{-10}$, a PAH$^-$ component; $x_{pah0}=1\times
10^{-8}$, and the corresponding PAH$^+$ and ion component;
$x_{i0}=1.1\times 10^{-8}$, then by charge neutrality,
$x_{e0}=6.606\times 10^{-10}$ (consistent with $x_{e0}$ in
\citet{KN96}). Since $x_{e0}$ is two orders of magnitude lower than
in the previous models (Sections
\ref{singlesizedgrainmodel}-\ref{MRNmantles}), there are not enough
electrons to charge all grain classes  inside the shock. Even if
only the smaller grain size classes are charged, which account for a
significant fraction of the total grain surface area, these are
already highly coupled to ${\bf B}$ ($|\beta_g|>>1$) and increasing
$|Z_g|$ does not effect the grain dynamics  or the shock structure.
For simplicity, $Z_g$ is   assumed constant  ($Z_g=-1$). The rate
coefficient for PAH-neutral scattering is the same as for the ions
with $<\sigma v>_{pah}=1.6 \times 10^{-9}\mathrm{cm^3s^{-1}}$
\citep{W98}, and $\beta_{pah0}=-\beta_{i0}= -1.382\times 10^4$.

The  shock profiles  ($\theta=45^\circ$) for MRN(PAH) and MRN are compared in Fig. \ref{mrntheta45withPAH}.
There is significantly more ${\bf B_\perp}$
rotation for the MRN(PAH)  model. The $B_x$ and   $T$ profiles are
similar, leading to  the  close similarities in $v_x$ $(v_{ix}-v_x)$ and $v_z$ $(v_{iz}-v_z)$.
Ions, electrons, and PAHs, remain well coupled  to ${\bf B}$ ,  and
$({\bf v}_i-{\bf v})=({\bf v}_e-{\bf v})=({\bf v}_{pah}-{\bf v})$.

There is a clear distinction between $(v_{gx,z}-v_{x,z})$ for the
smaller and larger grain classes. The smaller grains remain well coupled to ${\bf B}$,
whereas, the larger grains remain highly coupled to the
neutrals through collisions and are  co-moving with them
($v_{gx,z}-v_{x,z}\sim0$). With fixed  $Z_g$, the larger grains do
not become increasingly coupled  to ${\bf B}$ as they pass through
the shock (as in the previous results for MRN). From upstream to
downstream, each smaller grain is compressed ahead of the next
larger grain. The smallest grains  are compressed with the ions and
magnetic field through the entire shock, and the larger grains are
compressed with the neutrals. The midrange size class
ratios $v_s/v_{zg}$  lie between these two extremes.

 $(v_{gy}-v_y)$ show  (in general)  similar relationships between the relative
magnitude and sign with grain class size,  as seen previously in
Fig. \ref{MRNtheta45plot}. Smaller grain classes are coupled to
${\bf B}$  with $(v_{gy}-v_y)\sim(v_{iy}-v_y)>0$,
 and the larger grain classes have $(v_{gy}-v_y)<0$.
There is also a turnover in the magnitude of $(v_{gy}-v_y)$  with increasing $a_g$.
The larger grain size classes ($|\beta_g|<<1$) are highly coupled to the neutrals and  have small
 $|v_{gy}-v_y|$. As  $a_g$ decreases ($|\beta_g|\sim1-10$), Hall
effects are present and these obtain larger negative $(v_{gy}-v_y)$.
As $a_g$ is decreased further, the coupling to ${\bf B}$
increases, and  $(v_{gy}-v_y) \rightarrow (v_{iy}-v_y)$.


\subsection{Suppressing $B_y$}\label{sectionbysuppressed}

Fig. \ref{mrntheta45withPAHwithbysup} shows the effect of
suppressing $B_y$ in  the MRN(PAH) model.  $v_{x,y,z}$,
$(v_{gx,z}-v_{x,z})$ and $(v_{ix,z}-v_{x,z})$ are essentially
unaltered, except for the slightly higher peak in
 $(v_{ix}-v_x )$ for $B_y=0$, due to the
corresponding differences in $({\bf E'}\times{ \bf B})_x$.   The
peak in $T$ is lower for $B_y=0$, but only by $\sim 10\mathrm{K}$.
The heating rate is increased, however the cooling rate is a strong
function of $T$ and so the change in $T$ is small.   In $y$ all
grain size classes lag the neutrals ($v_{gy}-v_y)<0$, and for the
smallest grains $(v_{gy}-v_y)\sim0$. The larger grain size classes
have similar $(v_{gy}-v_y)$  as in Fig. \ref{mrntheta45withPAH} as
they are decoupled from ${\bf B}$, and so are unaffected by changes
in $B_y$. The corresponding $B_y=0$ results of the models in Sections \ref{singlesizedgrainmodel}-\ref{MRNwithPAHsection} plots not
shown)  also show only minor differences.

\section{Discussion}\label{discussion}

The results above demonstrate the effects of charged dust grains on
the structure of fast C-type shocks. For
$n_H=10^{5}\mathrm{cm}^{-3}$ and shock speed $v_s=18\mathrm{km/s}$
radiative C-type shock solutions exist $(\theta_{critical}<
\theta<90^\circ)$. The grain model details, i.e., modelling the
grain population as either a separate grain species or else as a
continuous grain size distribution, is important to the overall
shock structure, as well as the dynamics  of the individual grain
species/size classes.

The grain models considered here were chosen to illustrate the main
qualitative changes in  the resulting shocks. The shock profiles
were clearly altered with the introduction of a grain size
distribution as opposed to the single sized grain models. The net
charge on the grains is increased for the MRN models with the grains
becoming better coupled to the magnetic field, resulting in less
magnetic field rotation inside the shock front. The MRN(mantles) and
MRN(PAH) models were chosen to illustrate changes in the shock
profiles when departing from the standard MRN model.  It should be
noted that the MRN(mantles) model is not purely a MRN model with
growth of ice mantles of constant thickness. We have followed the
model of \cite{NNU91} who increase the size of the grains by a
factor of $9/5$. The qualitative changes are clear though; any
increase in the grain sizes leading to larger grain surface areas
(larger grain charge) will cause an increase in  the net charge on
the grains, and the grain coupling to ${\bf B}$. However, by
increasing the collisional cross section, the grains are
increasingly coupled to the neutrals via collisions. In the
MRN(mantles) models shown in section \ref{MRNmantles}, the latter
effect was dominant with the MRN(mantles) solution showing more
magnetic field rotation than it MRN counterpart.

As ${\bf B}$ is compressed  inside the shock,  the
charged species are accelerated with respect to the neutrals,
however, this is opposed by the collisional drag of the neutrals. If
the grains are highly coupled to ${\bf B}$ (high $|\beta_g|$) and the
movements of the neutrals and other highly coupled species (i.e.,
ions, PAHs, and electrons)  are mainly restricted to the $x-z$
plane, then $B_y$ is small. If however, the grains are only partially
coupled  to ${\bf B}$, either because they are large in
size  and so are better coupled to the neutral fluid via collisions
and/or have low charge $|Z_g|$ (and low $|\beta_g|$), then the
grains drift out of the $x-z$ plane, and the shock becomes
non-coplanar with large $B_y$. The grains  dominate the collisional heating of the neutrals, with
only a small contribution from the ions and a negligible component
from the electrons (due to their low mass).

The non-coplanarity in the shock solution depends on the grain
model; the single grain models (Sections
\ref{singlesizedgrainmodel}-\ref{singlesizedgrainmodelmantles})
produce significantly more rotation of ${\bf B}_\perp$ than for the
MRN  models (Sections \ref{MRNsection}-\ref{MRNwithPAHsection}). For
both $0.1\mu\mathrm{m}$ and $0.4\mu\mathrm{m}$ models, the grains
were only partially coupled to ${\bf B}$  with $|\beta_g|\sim
0.1-10$ inside the shock and the out-of-plane forces are
substantial. For the  MRN models, rotation of ${\bf B}$ is minimal,
since  the more abundant, smaller $a_g$ classes ($|\beta_g|>>1$)
which are highly coupled to ${\bf B}$, dominate the net grain
effects on the shock structure.

As $\theta \rightarrow 90^\circ$ the peak in  $B_y$ decreases, for a
given grain model, and  the well ${\bf B}$ coupled charged  species
are restricted to the shock plane (weakly coupled grains are still
able to drift out of the shock plane). ${\bf B}_\perp$ rotation
increases with decreasing $\theta$ and  $|{\bf v}_j -{\bf v}|$
become larger (on average), resulting in more dissipation. The shock
must dissipate the incoming kinetic energy flux over a shorter
length scale,   and the peak temperature $T$ inside low $\theta$
shocks (for a given grain size model) are therefore higher than for
high $\theta$ shocks. For low $\theta$ shocks, the neutral fluid
experiences more compression as it cools and so the sound speed is
higher  in the final part of the flow. There is a critical point for
low enough $\theta$, that in the neutral fluid $v_z=c_s$, in which
case the derivative $dP/dz$ is ill-defined (equation
(\ref{odepressure})) and the shock  solution is no longer C-type.
Below this critical angle $\theta<\theta_{critical}$  the shock
solutions may be either J-type or $\mathrm{C}^*$-type.
The limiting value of
$\theta_{critical}$ varied with grain model, but the ordering
remains same, the  higher the peak temperature in the shock, the
larger $\theta_{critical}$.

Grains dominate the collisional heating of the neutrals.  The peak
$T$  was higher and shock width narrower for the $0.1\mu\mathrm{m}$
model (Section \ref{singlesizedgrainmodel}) compared with the
$0.4\mu\mathrm{m}$ model (Section
\ref{singlesizedgrainmodelmantles}) since the pre-shock abundance of
the $0.1 \mu \mathrm{m}$ grains was two orders of magnitude larger,
than for the $0.4\mu{\mathrm{m}}$ model, which compensated for their
smaller collisional cross-sections. Conversely, for the MRN(mantles)
model (Section \ref{MRNmantles}), the increase in the total
frictional heating of the neutrals due to the increase in the net
grain collisional cross-sections  lead to hotter thinner shocks than
for the MRN model with an equivalent pre-shock grain abundance
(Section \ref{MRNsection}). Since the temperatures, and subsequently
$c_s$, are higher in the MRN(mantles) model (for a given $\theta$),
it is more difficult for the neutral fluid to maintain supersonic
flow and subsequently $\theta$ is restricted to
$\theta>\theta_{critical}\sim60^\circ$ for the MRN(mantles) model
but only $\theta>\theta_{critical}\sim45^\circ$ for the MRN model
(for $v_s$ and $n_H$ considered here).


There are  also critical values of $n_H$  for
which C-type solutions do not exist. Increasing  $n_H$ (for a given
$\theta$), increases the neutral-grain collisional heating rates and
thus dissipation.  With a large enough increase in $T$ and
$c_s$ the neutral flow cannot remain supersonic. For all grain
models considered here, the C-type solutions also broke down for
$n_H>10^5\mathrm{cm^{-3}}$ (independent of $\theta$),
with  only J-type or C$^*$type shocks possible. The exact limits
of $\theta$  and $n_H$ could vary with the choice  of cooling
function. The \citet{LS83} cooling rate has been used here, other
treatments of the cooling rate \citep{MT96,GP98},  will lead to
different temperature structures and thus different limits on
$\theta$ and $n_H$. However the orderings of the different  grain
models  will remain the same;  the cooler  MRN and
$0.4\mu\mathrm{m}$ models will have lower
$\theta_{critical}$ than the hotter MRN(mantles) and $0.1 \mu
\mathrm{m}$ models, respectively.

In the MRN models (Sections \ref{MRNsection}-\ref{MRNmantles}) the
$T_e$ dependent grain charging,  $Z_g(T_e)$, increased the grain
coupling to ${\bf B}$. The smaller $a_g$ classes  $|\beta_g|>>1$
were compressed with the ion fluid and ${\bf B}$. Mid range grains
remained coupled to ${\bf B}$ until reductions in $Z_g$ lead to
their decoupling from ${\bf B}$, and were compressed with the
neutral fluid thereafter. The largest grains $|\beta_g|<<1$ are
essentially compressed along with the neutral fluid. When PAHs are
present (Section \ref{MRNwithPAHsection}), $x_{e0}$  is reduced with
$Z_g=-1$, and  the mid to large grain size classes are less coupled
to ${\bf B}$  and are (on average) decoupled further by neutral
collisions inside the shock. As a result there is a larger Hall
current and the amount of rotation of ${\bf B}_\perp$ is increased.

Here $T_e$ is  given  by the approximation in  equation
(\ref{teapprox}) based upon Fig. 1 of \citet{DRD83}. Changing the
treatment of $T_e$ could largely effect $Z_g$  of the larger grains
and alter their coupling to ${\bf B}$  as well as $({\bf v}_g -{\bf
v})$. However, smaller grains with low $Z_g$  are highly coupled to
${\bf B}$ (essentially independent of  $T_e$ and $Z_g$) and dominate
the grain effects on shock structure. Thus variations in the
treatment of $T_e$ are unlikely to change the model results.


$|Z_g|$ increases with increasing $T_e$  according to equation
(\ref{graincharge}), however  as $T_e$ decreases, the uncharging
proceeds simply via  the re-release of the captured electrons. More
realistically  the reduction in $|Z_g|$ proceeds via the capture of
ions \citep{DS87}, and the ion fluid is depleted. Moreover, inside
shock waves in  dense clouds, negatively charged PAHs may be
neutralized in neutral-PAH$^-$ reactions, resulting in electron
detachment \citep{PEA88}. \citet{FP03} modelled the charge
variations of grains in a planar steady C-type shock with
$v_s=50\mathrm{km/s}$. The grain population was represented by  a
small (dominant)  PAH grain component and a large grain component
(given by an MRN distribution $0.01\le a_g \le0.3\mu \mathrm{m}$),
and they found that the charge variations for each were very
different. The PAHs  were neutralised by neutral-PAH$^-$ reactions
via electron detachment, while the larger grains became more
negatively charged as $T_e$ and thus $|Z_g|$ increased through the
shock.

Detachment of electrons via neutralization of the PAH$^-$s,
 may allow for the $Z_g(a,T_e)$ (equation (\ref{graincharge}))
 charging to proceed in the MRN(PAH) model, potentially
increasing the grain-${\bf B}$ coupling, reducing the rotation of ${\bf
B}_\perp$. However, the capture of ions could also counteract any
increases in $Z_g$. The shock model of \citet{FP03},  included both
electron and ion attachment on the grain surfaces, and the net
negative charge on the grain component still increased significantly
due to the increase in the attachment rate of electrons with
increasing $T_e$. Thus, it is likely that if ion attachment were
included in the models here,  $|Z_g|$ should still increase with
$T_e$. The details of this will need to be quantified and confirmed
in further studies.

Inertia of the grains has also been neglected and their drift is
given by the balance of electromagnetic forces and neutral-grain
collisional drag. \citet{CR02} demonstrated that for very small
grains  ($a_g<0.1\mu \mathrm{m}$) the length scales over which  the
neutral fluid and magnetic field were compressed, $L_{neut}$  and
$L_{field}$, respectively,  are over three orders of magnitude
larger than the length scales for the gas drag to decelerate the
grains $L_{drag}(g)$. Thus, neglecting the inertia of the smaller
grains is a good approximation. However, for the large grains ($a_g
\gapprox 0.1\mu \mathrm{m}$), $L_{drag}(g)$ is of the same order as
$L_{field}$ and so the inertia cannot be neglected. The larger
grains dominate the field-neutral coupling  and so the inclusion of
their inertia may effect the shock structure.


The amount of dissipation inside the shock was  relatively
insensitive to the orientation of ${\bf B}_0$ for the pre-shock
conditions considered here; $n_H=10^5\mathrm{cm}^{-3}$. The  peak
$T$ inside the shock was essentially unaltered  when $B_y =0$ for
the MRN(PAH) model (Section \ref{sectionbysuppressed}) even though
the MRN(PAH) model had the most ${\bf B}_\perp$ rotation. Similar
comparisons have been made for the other MRN models. The critical
angle $\theta_{critical}$ for which the neutral flow inside the
shock makes the transition from supersonic to subsonic is also
unaltered when $B_y$ is suppressed. For  the purposes of calculating
detailed chemical models and molecular line emission the shock
problem can be greatly simplified by suppressing $B_y$.
 This may not be the case, however, for higher
$n_H$ where grains may become further decoupled from ${\bf B}$.

The choice of grain model affects the shock structure, however, the
same overall jump conditions (for given pre-shock
conditions) are still satisfied, and the same power per unit area  is
dissipated inside the layer of hot gas inside the shock transition.
For the models here, the temperatures inside the shock reach
$\sim 1500\mathrm{K}$ and  the molecules will remain intact, so the
molecular line emission should be similar regardless of the grain
model. \citet{NS97} numerically modelled CO and H$_2$ line
emission for C-type shocks subject to the Wardle instability,
concluding that the line strengths are relatively insensitive to the exact
details of the shock structure.   However, the choice
of grain model does effect the limits
$n_H$ and $\theta$ for which the shock becomes J-type (or C$^*$
type), and the temperatures inside J-type shock reach much higher
values, so the shocked molecules may not always remain intact.

\section{Summary}\label{summary}
New shock models for fast oblique C-type shocks have been presented
here with pre-shock conditions $n_H=10^5\mathrm{cm}^{-3}$, $v_s = 18
\mathrm{km/s}$, and  $M_A=10$ for freely specified magnetic field
${\bf B}_0$  orientation $35^\circ \le \theta \le 80^\circ$. The
grain population is represented by either
\begin{itemize}
\item a single grain population of size $0.1\mu\mathrm{m}$ or
$0.4\mu\mathrm{m}$,
\item a continuous (standard) MRN grain size distribution,
\item a MRN distribution with  ice mantles, or
\item a MRN distribution with additional populations of positively and negatively charged
PAHs.
\end{itemize}
There were clear differences in the shock profiles with grain model
due to the  coupling of the grains to either/both the magnetic field
and/or neutral fluid. Smaller and/or highly charged grains remain
well coupled to ${\bf B}$ whereas larger grains are coupled to the
neutral fluid through collisions. Shock models with grain
populations that are well coupled to  ${\bf B}$ result in more
co-planar shock profiles with minimal rotation of ${\bf B}$  out of
the shock plane. The results are summarized below:
\begin{itemize}
\item For a given $\theta$, the smaller $0.1\mu\mathrm{m}$ grain model produces more
co-planar solutions than the $0.4\mu\mathrm{m}$ model since the
$0.1\mu\mathrm{m}$ grains are better coupled to ${\bf B}$.
\item Each grain size class in the MRN models were
differentially accelerated inside the shock front, with the largest
grains being compressed along with the neutrals and the smaller
grains compressed along with the ions and ${\bf B}$.
\item The more abundant, smaller grains in the MRN distribution models dominate the
net grain effects on the shock structure resulting in more co-planar
solutions compared with the single grain models.
\item The effect of charging  grains via the capture of the
electrons increased the grain-${\bf B}$ coupling in all grain
models.
\item The increase in the grain surface areas from the MRN to the
MRN(mantles) models increased the net grain charge, however the
increase in the grain collisional cross sections resulted in better
coupling of the grains to the neutrals and consequently more
rotation of ${\bf B}$ out of the shock plane, for a given $\theta$.
\item  The grains dominate the drag force and collisional heating of the
neutrals, so  the models with larger (larger collisional cross
sections) or more abundant grains  have higher peak temperatures and
the shock fronts are subsequently narrower i.e., the collisional
heating  for the less abundant $0.4\mu\mathrm{m}$ grains lead to
lower peak shock temperatures and wider shock fronts  than for the
more abundant $0.1\mu\mathrm{m}$ grains. Conversely, the larger
MRN(mantles) model lead to increased peak temperatures, and narrower
shock fronts, compared with the MRN model, due to the increase in
the collisional cross section of the grains.

\item The higher the peak temperatures reached inside the shock
front, the higher the sound speed and the neutrals cannot always
remain supersonic. As a result, the lower limit of $\theta$ for
which C-type solutions are possible is dependent on grain model;
$\theta_{critical}\sim40^\circ$ for the $0.1 \mu\mathrm{m}$ model,
$\theta_{critical}\sim45^\circ$ for the MRN and MRN(PAHs) models,
$\theta_{critical}\sim60^\circ$ for the MRN(mantles) model (for
$v_s$ and $n_H$ considered here).

\item The non-coplanarity of the shock solution is also dependent
upon the orientation of ${\bf B}_0$, with the peak in $B_y$
decreasing as $\theta\rightarrow 90^\circ$.

\item For pre-shock conditions considered here
($n_H=10^5\mathrm{cm}^{-3}$, $v_s = 18 \mathrm{km/s}$) the amount of
dissipation inside the shock is relatively insensitive to the
orientation of ${\bf B}_0$, so in calculating  molecular line
emission or chemical models the shock problem is greatly simplified
by suppressing $B_y$.

\end{itemize}

\section*{Acknowledgments}
This research was supported by the Australian Research Council.

\appendix

\label{lastpage}

\end{document}